%% file: main.tex
\RequirePackage{silence}
\documentclass[%
reprint,
superscriptaddress,
aps,
prb,
floatfix,
showkeys,
]{revtex4-2}

\input{Organisation/packages}
\input{Organisation/shorthand}

\begin{document}

\title{Theory of Optical and Magneto-Optical Properties of [111]-Oriented Site-Controlled Quantum Dots}

\author{N. D. O'Connor}
 \email{noconnor@umail.ucc.ie}
 \affiliation{%
 Tyndall National Institute, University College Cork, Cork, Ireland.
 }
 \affiliation{
 School of Physics University College Cork, Cork, Ireland.
 }
\author{G. Ju\v{s}ka}
 \affiliation{%
 Tyndall National Institute, University College Cork, Cork, Ireland.
 }
\author{E. Pelucchi}
 \affiliation{%
 Tyndall National Institute, University College Cork, Cork, Ireland.
 }
\author{S. Schulz}%
 \email{stefan.schulz@tyndall.ie}
 \affiliation{%
 Tyndall National Institute, University College Cork, Cork, Ireland.
 }
 \affiliation{
 School of Physics University College Cork, Cork, Ireland.
 }

\date{\today}

\begin{abstract}
Site-controlled [111]-oriented quantum dots (QDs) are promising sources of non-classical light, yet their optical spectra exhibit apparently conflicting symmetry signatures. At zero magnetic field, excitonic spectra are commonly interpreted as \sym{C}{2v}-like, whereas measurements in Faraday geometry display selection rules characteristic of \sym{C}{3v}-like symmetry. Here, we show that both regimes can be described consistently without invoking any change in the underlying structural symmetry. Combining 8-band \kp theory, a self-consistent Hartree-informed configuration interaction scheme, and group theoretical analysis, we introduce a controlled symmetry reducing perturbation that lowers the QD symmetry to \sym{C}{s}. At zero field, the optical response remains predominantly \sym{C}{2v}-like because the additional low-symmetry transitions carry only weak oscillator strengths. In Faraday geometry, the same reduced-symmetry structure produces four dominant, nearly circularly polarised trion transitions, reproducing the experimentally observed \sym{C}{3v}-like response. By contrast, an exact \sym{C}{2v} description inferred from the zero-field spectrum permits only two such trion transitions and therefore fails at finite field. Our results thus provide a unified description of the zero and finite magnetic field optical response without any restoration or enhancement of the underlying structural symmetry. More generally, we demonstrate that a higher-symmetry-like optical spectrum does not uniquely identify the exact symmetry of a QD, therefore also providing useful guidance for other QD and nanostructure systems that lie beyond the ones studied here.
\end{abstract}

\maketitle

\begin{bibunit}

\input{sections/1_Intro/introduction}
\input{sections/1_Intro/exp_overview}
\input{sections/2_Theory/theo_frame}
\input{sections/3_Results/FSS_anisotropy}
\input{sections/3_Results/magnetic}
\input{sections/4_Conclusion/conclusion}
\input{sections/5_Acknowledgements/acknowledgement}

\putbib

\end{bibunit}

\clearpage

\widetext

\begin{bibunit}

\input{Organisation/reset_SI}
\input{appendices/numerics}
\input{appendices/GT}
\input{appendices/OMEs}

\putbib

\end{bibunit}

\end{document}

%% file: Organisation/packages.tex
\usepackage[english]{babel}
\usepackage[T1]{fontenc}
\usepackage{amsmath,amssymb,amsfonts,mathrsfs}
\usepackage{bm}
\usepackage{braket}
\usepackage{booktabs}
\usepackage{cellspace}
\usepackage{soul}
\usepackage{dcolumn}
\usepackage{graphicx}
\usepackage{hyperref}
\usepackage[utf8]{inputenc}
\usepackage{stfloats}
\usepackage{textcomp}
\usepackage{xfrac}
\usepackage{lmodern}
\usepackage[table]{xcolor}
\usepackage[normalem]{ulem}
\usepackage{bibunits}
\defaultbibliographystyle{apsrev4-2}
\defaultbibliography{Organisation/references}

\makeatletter
\AddToHook{begindocument/end}{%

    \let\@bibsetup\NATx@bibsetnum

    \def\endthebibliography{%
        \auto@bib@innerbib
        \endNAT@thebibliography
        \aftergroup\auto@bib@empty
    }%
}
\makeatother

\makeatletter
\def\SI@toc{toc}
\newcommand\SIstartcontents{%
  \let\SIorigaddcontentsline\addcontentsline
  \renewcommand\addcontentsline[1]{%
    \def\SI@ext{##1}%
    \ifx\SI@ext\SI@toc\def\SI@ext{stc}\fi
    \expandafter\SIorigaddcontentsline\expandafter{\SI@ext}}}
\newcommand\SItableofcontents{\@starttoc{stc}}
\makeatother

%% file: Organisation/shorthand.tex
\usepackage{xspace}
\newcommand{\kp}{$\mathbf{k}\cdot\mathbf{p}$\xspace}
\newcommand{\sym}[2]{\ensuremath{\mathrm{#1}_{\mathrm{#2}}}\xspace}
\newcommand{\pe}{\ensuremath{\psi_{e,0}}\xspace}
\newcommand{\phz}{\ensuremath{\psi_{h,0}}\xspace}
\newcommand{\pho}{\ensuremath{\psi_{h,1}}\xspace}
\newcommand{\pht}{\ensuremath{\psi_{h,2}}\xspace}

%% file: sections/1_Intro/introduction.tex
\section{\label{sec:Intro} Introduction}

Semiconductor quantum dots (QDs) are promising candidates for on-demand non-classical light emission due to their discrete, atomic-like energy levels arising from three dimensional quantum confinement~\cite{TrFi2020}. In particular, the biexciton-exciton (XX-X) radiative recombination cascade has attracted significant attention as a potential route towards deterministic generation of polarisation entangled photon pairs~\cite{BeSa2000}. The fidelity of the emitted entangled state is, however, strongly influenced by the magnitude of the so-called excitonic fine structure splitting (FSS), which describes the degeneracy lifting of the bright exciton states in the radiative cascade~\cite{SiBe2010}. Understanding and controlling structural and electronic properties that determine the FSS has therefore become a central aim in the development of QDs for entangled photon generation via the biexciton-exciton cascade.

In general, the FSS is closely linked to the symmetry of the carrier confinement potential. Conventional [001]-oriented zincblende QDs, which are widely studied dot systems for entangled photon emission ~\cite{YoSt2006,AkLi2006a,BaSa2026,PiRo2026}, are in general described by a \sym{C}{2v} point group symmetry~\cite{BaOr2002,ScWi2007}. Such a symmetry does not protect the degeneracy of the bright exciton pair and fundamentally leads to a non-vanishing FSS~\cite{BeNa2003}. Therefore, various approaches have been developed to reduce or eliminate the FSS by, e.g. applying external fields~\cite{HuRe2018a,PoBe2014, WaGo2012}. On the other hand, growth of [111]-oriented zincblende QDs has been proposed as a further solution to this problem, given that these structures ideally exhibit a \sym{C}{3v} point group symmetry, which entails a vanishing excitonic FSS~\cite{ScWi2009a, KaDu2010, ScCa2011, SiBe2009}. Experimental studies, using site-controlled [111]-oriented pyramidal (In,Ga)As QDs, have indeed confirmed a near vanishing FSS and thus their suitability for entangled photon emission~\cite{JuDi2013,PiBa2023,KuMa2013}. However, it is worth noting that in order to explain the experimentally observed optical signatures, a proposed \sym{C}{2v}-like symmetry description has to be invoked~\cite{KaDu2010}. Nevertheless, the FSS values found experimentally are small and reproducible~\cite{JuMu2015, RaJu2021a} without employing external fields to tune and tailor their excitonic properties.

In addition to their highly attractive excitonic properties, site-controlled growth enables deterministic spatial positioning of individual QDs together with a high degree of control over structural properties, such as dot dimensions, alloy composition and nanostructures' stacking~\cite{RaJu2021a, MeDi2009, ZhKa2009, MoCh2017}. Such control is important not only for polarisation entangled photon sources but also for more advanced routes to non-classical light, including emerging photonic cluster state generation protocols~\cite{LiRu2009,EcLi2010, VeHi2022,ScCo2016,CoSu2023}.

More specifically, these photonic cluster state schemes typically rely on a finite external magnetic field to control the QDs electronic and excitonic properties~\cite{EcLi2010, VeHi2022, CoSu2023}. Key for these protocols are the magneto-optical selection rules, which are determined by the combined symmetry of QD geometry and the direction of the applied magnetic field~\cite{BrCr2010}. Previous theoretical and experimental studies of [111]-oriented (In,Ga)As QDs revealed that a \sym{C}{3v}-like symmetry description of the optical selection rules needs to be applied to explain the (multi-)excitonic emission spectra in the presence of a magnetic field along the [111] crystallographic axis~\cite{BaSa2025a, BaSa2026b, DuGl2013}, i.e., in so-called Faraday configuration. From a group theoretical perspective the Faraday geometry trion spectrum of a \sym{C}{3v} symmetric QD is markedly different from that expected for \sym{C}{2v} symmetry as already highlighted in the literature~\cite{DuGl2013}. This therefore gives rise to an apparent contradiction for [111]-oriented site-controlled (In,Ga)As QDs: in the absence of a magnetic field, a \sym{C}{2v}-like description of the optical response is required to explain the excitonic spectra, e.g. the FSS, whereas with a magnetic field along the [111]-direction, the system instead exhibits a \sym{C}{3v}-like optical response.

In this work, we address the apparently conflicting symmetry signatures reported for [111]-oriented site-controlled QDs at zero and finite magnetic fields. We combine continuum 8-band \kp theory, self-consistent Hartree informed configuration interaction (CI), and group theory to determine how symmetry influences the electronic structure and optical selection rules of [111]-oriented arsenide QDs. In particular, we introduce a controlled geometrical perturbation that reduces the symmetry of the full zero-field Hamiltonian to \sym{C}{s} and investigate whether this underlying reduced symmetry can provide a unified description of the optical properties in both the absence and presence of an external magnetic field. 

We show that the answer lies in distinguishing the exact symmetry of the Hamiltonian from the symmetry character of the dominant optical response. At zero field, the exact \sym{C}{s}-symmetry system exhibits a predominantly \sym{C}{2v}-like optical response because the additional transitions allowed by the lower symmetry remain strongly suppressed. In Faraday geometry, the underlying reduced-symmetry structure remains unchanged, yet the dominant trion response closely resembles that expected for the higher \sym{C}{3v} geometric symmetry reference. The \sym{C}{2v}-like zero-field description therefore cannot simply be carried into a finite magnetic field analysis.

Our results provide a unified and consistent description of the apparently conflicting optical symmetry signatures observed at zero and finite magnetic fields of [111]-oriented arsenide based QDs, without requiring any restoration or enhancement of the underlying structural symmetry. More generally, we demonstrate that optical selection-rule patterns characteristic of a higher-symmetry reference group do not necessarily reflect the exact symmetry of the underlying QD Hamiltonian. This finding has broader implications for the interpretation of optical and magneto-optical spectra in semiconductor QDs and other low-dimensional nanostructures in general, highlighting the need to distinguish between the exact symmetry of a system and the apparent symmetry inferred from its optical response.

The manuscript is structured as follows. Section~\ref{sec:ExpOver} summarizes briefly previous experimental and theoretical studies, allowing us to establish the generally observed zero and finite magnetic field behaviour of site-controlled [111]-oriented QDs. In Sec.~\ref{sec:Num_Mod} the \kp, self-consistent Hartree and CI theoretical framework is outlined. Section~\ref{sec:Res1} examines the zero-field electronic and excitonic structure, introducing the relevant selection rules along with calculated optical spectra. The electronic and optical properties of the studied QDs under the influence of a magnetic field in Faraday configuration  are presented in Sec.~\ref{sec:ResMag}, including a discussion of the corresponding magnetic group selection rules.

%% file: sections/1_Intro/exp_overview.tex
\section{\label{sec:ExpOver} Previous Studies of Symmetry and Optical Selection Rules in [111]-Oriented Arsenide Quantum Dots}

Site-controlled pyramidal QDs grown along the [111]-direction of the zincblende lattice have been investigated as a route to highly uniform emitters with a reproducibly very small excitonic FSS~\cite{KaOb2015, BaWa2004, HeYo2010, MeMa2012, HaDu1999}. Early measurements of site-controlled pyramidal (In,Ga)As QDs indeed demonstrated a high degree of spectral reproducibility together with relatively small, but generally non-zero, FSS values, with a mean value of approximately $2.1 \pm 1.2\,\mu \mathrm{eV}$ reported for one particular ensemble~\cite{JuMu2015}. More recently, Ranjbar Jahromi \textit{et al.} investigated site-controlled GaAs/(Al,Ga)As pyramidal QDs and demonstrated strategies by which the average FSS could be reduced to values of only a few $\mu\text{eV}$~\cite{RaJu2021a}. 

A detailed symmetry analysis of the excitonic properties of [111]-oriented pyramidal QDs was presented by Karlsson \emph{et al.}~\cite{KaDu2010}. Combining polarisation-resolved photoluminescence (PL) spectroscopy with numerical modelling and group-theoretical arguments, they extensively investigated the fine structure and optical selection rules of multi-excitonic complexes in these QD structures. Their analysis showed that the experimental spectra could not be solely described in terms of ideal \sym{C}{3v} symmetry. Instead, the optical response exhibited signatures of both weak symmetry breaking and an additional approximate symmetry plane, motivating the use of reduced- and elevated-symmetry descriptions for different parts of the excitonic spectrum, with the authors concluding that in the absence of a magnetic field, a \sym{C}{2v}-like description provided good agreement with the experimental spectrum. Importantly, all measurements, modelling, and symmetry arguments in Ref.~\cite{KaDu2010} were carried out in the absence of an applied magnetic field.

The effect of an applied magnetic field on the electronic and optical properties of [111]-oriented QDs was investigated experimentally and theoretically by Durnev \emph{et al.}~\cite{DuGl2013}. This study considered [111]-oriented GaAs/(Al,Ga)As droplet QDs. While the growth process is slightly different to that of the site-controlled pyramidal structures discussed above, the work by Durnev \emph{et al.}~\cite{DuGl2013} provides an important reference for the magneto-optical response expected from a geometrically \sym{C}{3v} symmetric QD. For a geometrically \sym{C}{3v} system, a longitudinal magnetic field (Faraday geometry), $B\parallel[111]$ aligned with the \sym{C}{3} rotation axis, can mix the total angular momentum components, $\ket{J,m_j}$, of a heavy-hole (HH), $\ket{3/2, +3/2}$ and $\ket{3/2, -3/2}$ basis states, respectively. For positively charged excitons (trions), this HH mixing gives rise to four characteristic bright transitions in Faraday geometry, comprising two $\sigma^{+}$- and two $\sigma^{-}$-polarised lines. This is in clear contrast to the \sym{C}{2v} reference case, relevant for conventional [001]-oriented SK QDs, where only two circularly polarised transitions, one of each helicity, are symmetry allowed~\cite{BaOr2002,DuGl2013,BaGe2011}.

Experimentally, Durnev \emph{et al.}~\cite{DuGl2013} observed the four-line pattern expected for \sym{C}{3v} symmetry and inconsistent with \sym{C}{2v}. Yet, in the same GaAs/(Al,Ga)As [111] droplet QD system, the zero-field spectra exhibit a finite, although strongly suppressed, FSS~\cite{MaAb2010}, indicating a departure from ideal \sym{C}{3v} symmetry. Thus, the same QD system appears \sym{C}{2v}-like at zero field but \sym{C}{3v}-like in a Faraday geometry magnetic field. More recently, Barcan \emph{et al.}~\cite{BaSa2025a, BaSa2026b} likewise observed a four transition trion response in site-controlled (In,Ga)As QDs in Faraday geometry. These results show that the finite field optical behaviour can closely resemble that of an ideal \sym{C}{3v} geometric reference, despite the lower symmetry (\sym{C}{2v}-like excitonic FSS) optical signatures observed at zero field.

Taken together, the literature therefore presents an apparent contrast, at zero magnetic field, the optical response of site-controlled [111]-oriented QDs has been successfully described as \sym{C}{2v}-like~\cite{KaDu2010}, whereas in Faraday geometry the dominant trion response suggests a higher \sym{C}{3v} geometric description~\cite{ObBy2009, Ob2012}. Whether both signatures can arise from the same underlying reduced symmetry QD structure is the central question addressed below.

%% file: sections/2_Theory/theo_frame.tex
\section{\label{sec:Num_Mod} Theoretical Framework}

To gain insight into the electronic and multi-excitonic properties of [111]-oriented arsenide-based QDs, and to understand the factors driving the symmetry of such a system in the presence and absence of magnetic fields, we employ the following theoretical modelling framework. We begin by calculating the single particle states using a numerically efficient, three dimensional 8-band \kp model, that accounts for strain, piezoelectric and magnetic fields. To study multi-excitonic states, we employ a self-consistent Hartree informed many-body CI scheme. Our numerical analysis is accompanied by group theoretical considerations. 

\subsection{\label{subsec:SP}Single Particle Hamiltonian}

The QD single particle states are calculated using a three-dimensional, 8-band \kp envelope function approximation, implemented via finite differences. The Hamiltonian is expressed in the total angular momentum basis, $\ket{J, m_{j}}$, with $z$ parallel to the $[111]$ crystallographic direction. The conduction band basis states are $\ket{1/2,\pm1/2}_c$, while the valence band states $\ket{3/2,\pm3/2}_v$, $\ket{3/2,\pm1/2}_v$, and $\ket{1/2,\pm1/2}_v$ are to the heavy-hole (HH), light-hole (LH) and spin-orbit split-off (SO) basis states, respectively. This Hamiltonian was obtained by rotating the conventional [001]-oriented Hamiltonian from Ref.~\cite{Wi2003}, following the general procedure described in Refs.~\cite{WiLe2009, MaO2014}. The total Hamiltonian is written as
\begin{equation}
    H_{\mathbf{k}\cdot\mathbf{p}} = H_{\text{pot}} + H_{\text{kin}} + H_{\text{str}} + H_{\text{Z}} + V_p\,.
\end{equation}
Here, $H_{\text{pot}}$ contains the confinement (band-edge potential) terms describing the energy landscape of the system. Kinetic-energy terms up to second order in $\mathbf{k}$ are described by $H_{\text{kin}}$, while $H_{\text{str}}$ contains the strain dependent deformation potential terms and $V_p$ is the electrostatic potential energy arising from first- and second-order piezoelectric polarisation fields~\cite{ScCa2011}. The orbital effects of an external magnetic field are incorporated into $H_{\text{kin}}$ through the gauge-covariant derivative, using the gauge-invariant discretisation of Ref.~\cite{AnMo2008}. The spin Zeeman interaction is contained in $H_{\text{Z}}$, including remote-band contributions parametrised by $g'$, $\kappa$, and $q$. For the Zeeman term, we adopt the formulation of Ref.~\cite{MiGa2018}, rotated here to the [111] growth direction.

To ensure numerical stability and suppress spurious solutions often encountered in 8-band \kp models, we employ the zone-centre effective-mass rescaling of Refs.~\cite{AnSe2010,VeSt2007}, together with a specialised piecewise derivative scheme. Full details of the finite-difference implementation, along with all relevant material parameters are provided in the supplementary information. Diagonalisation of $H_{\mathbf{k}\cdot\mathbf{p}}$ yields the single-particle eigenenergies $\varepsilon_{\alpha,i}$ and wave functions $\psi_{\alpha,i}$, where $\alpha\in\{e,h\}$ labels the carrier species and $i$ denotes the single-particle state index. The resulting single-particle states and energies can now form a basis for the multi-excitonic calculations described in the following section.

\subsection{Many-Body Configuration Interaction}

To obtain multi-excitonic states, a configuration interaction (CI) scheme is employed. In the CI framework, the many-body wave functions are expanded, in general, in a basis of Slater determinants constructed from single-particle electron and hole states. In doing so, the CI approach treats the Coulomb interaction exactly within the many-body Hilbert space spanned by the chosen set of single-particle states. The 
basis-set truncation can be systematically controlled by increasing the number of single-particle states, and hence the number of Slater determinants, included in the CI expansion. The CI Hamiltonian in its general form reads:
\begin{equation}
    H_{\text{CI}} = H_e + H_h + V_{ee} + V_{hh} + V_{eh}^{dir} + V_{eh}^{ex}\,\, .
    \label{eq:CI_Ham}
\end{equation}
Here, $H_e$ and $H_h$ are the electron and hole one-particle operators, $V_{ee}$ and $V_{hh}$ describe electron-electron and hole-hole repulsion, while $V_{eh}^{dir}$ and $V_{eh}^{ex}$ denote the direct and exchange electron-hole interactions, respectively. In the following the CI basis expansion is restricted to single particle states bound to the QD, to avoid numerical artifacts in calculations including magnetic fields~\cite{ScGa2025}.

To incorporate mean-field Coulomb effects already at the single-particle level, the CI expansion is not constructed directly from the eigenstates of $H_{\mathbf{k}\cdot\mathbf{p}}$. Instead, first, a self-consistent Hartree calculation is performed. This procedure yields Coulomb-renormalised single-particle states $\tilde{\psi}_{\alpha,i}$ and energies $\tilde{\varepsilon}_{\alpha,i}$, which are then used to construct the CI basis. By incorporating the dominant mean-field Coulomb response already at the single-particle level, the Hartree procedure provides a basis that is better adapted to the interacting system and can substantially accelerate the convergence of the CI expansion~\cite{AbHa2006}. This is particularly advantageous when the CI basis is restricted to a relatively small number of QD-bound single-particle states.

The Coulomb matrix elements (CMEs) required for constructing the CI Hamiltonian, Eq.~\eqref{eq:CI_Ham}, are evaluated using $\tilde{\psi}_{\alpha,i}$. The CI Hamiltonian in second quantization, details of the self-consistent Hartree procedure, including correction for double-counting, and the numerical evaluation of the CMEs are provided in the supplementary information. Diagonalisation of $H_{\text{CI}}$ then yields the correlated many-body states, $\Psi_\alpha$, and energies, $E_\alpha$.

\subsection{Optical Spectra}
To calculate polarisation resolved optical spectra and thus to facilitate comparison with experiment and group theoretical selection rules, we use Fermi's golden rule~\cite{Di1978, LaLi2007}. Since the polarization characteristics of photons emitted from [111]-oriented QDs are central to distinguishing the optical selection rules associated with different symmetry classes, we obtain optical spectra from
\begin{equation}
    I^{\hat{\mathbf{e}}}_{if} \propto |\bra{\Psi_f}\hat{\mathbf{e}}\cdot\mathbf{D}\ket{\Psi_i}|^2\,\delta(E_i - E_f -\hbar\omega)\,.
\end{equation}
Here, $|\Psi_i\rangle$ and $|\Psi_f\rangle$ denote the initial and final multi-excitonic states of an optical transition, facilitated by the dipole operator $\mathbf{D}$ with light polarisation characterised by the vector $\hat{\mathbf{e}}$.

\input{FIGURES/1_vPol_Geom/figure}

In second quantization, the dipole operator is written as $\mathbf{D} = \sum_{\alpha,\beta} \mathbf{d}_{\alpha\beta}\, e_\alpha h_\beta+h.c.$~\cite{BaSc2007}. Given that the many-body states $|\Psi_i\rangle$ and $|\Psi_f\rangle$ are expressed as linear combinations of, in our case, mean-field corrected single-particle states, we calculate $\mathbf{d}_{\alpha\beta}$ via~\cite{St2005},
\begin{equation}
    \mathbf{d}_{\alpha\beta}=-\frac{i}{\tilde\varepsilon_{e,\alpha}-\tilde\varepsilon_{h,\beta}}\left\langle\tilde\psi_{e,\alpha}\middle|
    \left.\frac{\partial H_{\mathbf{k}\cdot\mathbf{p}}}{\partial \mathbf{k}}\right|_{\mathbf{k}=\mathbf{0}}\middle|\tilde\psi_{h,\beta}\right\rangle\,,
    \label{eq:HFT}
\end{equation}
where $|\tilde\psi_{e,\alpha}\rangle$ is an electron and $|\tilde\psi_{h,\beta}\rangle$ a hole state; $\tilde\varepsilon_{e,\alpha}$ and $\tilde\varepsilon_{h,\beta}$ are their corresponding energies. The momentum operator $\frac{\partial H_{\mathbf{k}\cdot\mathbf{p}}}{\partial \mathbf{k}}$ is the vector of directional derivatives along the two in-plane and one out-of-plane directions, so that $\mathbf{d}_{\alpha\beta} = (d^x_{\alpha\beta}, d^y_{\alpha\beta}, d^z_{\alpha\beta})$ carries information about the polarisation of an optical transition. For the light polarisation vector, we consider in the following $\hat{\mathbf{e}}_x=\hat{x}=\frac{1}{\sqrt{6}}[1,1,\bar{2}]$, $\hat{\mathbf{e}}_y=\hat{y}=\frac{1}{\sqrt{2}}[\bar{1},1,0]$, $\hat{\mathbf{e}}_z=\hat{z}=\frac{1}{\sqrt{3}}[1,1,1]$ and circularly polarized light ($\hat{\sigma^{\pm}}=\frac{\hat{x}\pm i\hat{y}}{\sqrt{2}}=\frac{\mathbf{e}_x\pm i\mathbf{e}_y}{\sqrt{2}}$) for comparison with experiment and group theoretical predictions for different symmetry groups.

%% file: FIGURES/1_vPol_Geom/figure.tex
\begin{figure}[t]

    \includegraphics[width=\columnwidth]{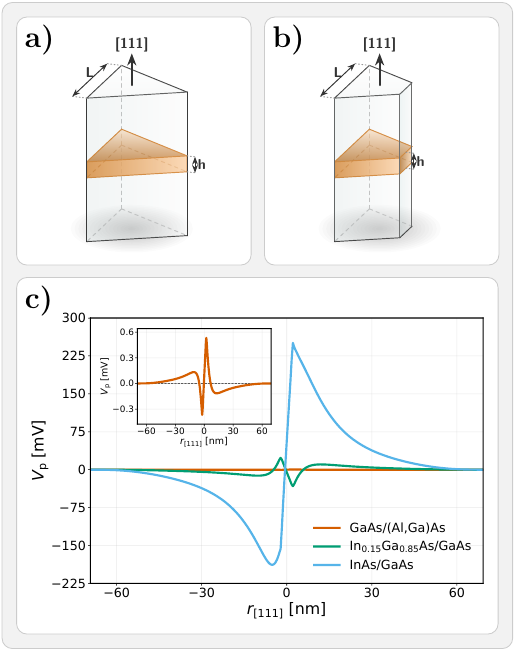}

    \caption{a) Schematic illustration of the ideal triangular quantum dot (QD) confinement geometry (orange) embedded in the vertical quantum wire (VQWr). The QD has lateral size $L = 26$~nm and height $h = 3$~nm grown along $[111]$ direction. The ideal confinement geometry possesses a \sym{D}{3h} symmetry. b) Corresponding structure with one corner of both the QD and VQWr truncated, reducing the geometric symmetry to \sym{C}{2v}. c) Comparison of the polarisation potential, $V_p$, along the [111] growth direction through the centre of GaAs/(Al,Ga)As, In$_{0.15}$Ga$_{0.85}$As/GaAs and InAs/GaAs QDs. The VQWr is only included for the GaAs/(Al,Ga)As system, while the In$_{0.15}$Ga$_{0.85}$As/GaAs and InAs/GaAs calculations use isolated QDs of the same nominal dimensions. Inset: Zoomed-in view of the GaAs/(Al,Ga)As potential.}
    \label{fig:vpol_111_comparison}
\end{figure}

%% file: sections/3_Results/FSS_anisotropy.tex
\section{\label{sec:Res1}Zero Magnetic Field Electronic and Excitonic Structure}

In this section, we consider the zero-field electronic and excitonic properties of site-controlled [111]-oriented GaAs/(Al,Ga)As QDs. The weak lattice mismatch and the absence of alloy disorder within the GaAs QD make them an ideal platform for isolating symmetry-reducing effects. Experimental studies on GaAs/(Al,Ga)As QDs have revealed small but finite excitonic FSSs~\cite{RaJu2021a}, indicating a departure from the ideal geometrical \sym{C}{3v} symmetry. Atomistic calculations by Pirard \emph{et al.}~\cite{PiBa2023} showed that random alloy fluctuations in the surrounding (Al,Ga)As barrier can induce a finite excitonic FSS. For the structures considered there, alloy disorder alone produced only relatively small, $\approx2\,\mathrm{{ \mu eV}}$, splittings. Pirard \emph{et al.}~\cite{PiBa2023} demonstrated that structural anisotropies, such as in-plane elongation of [111]-oriented GaAs QDs, dominate the FSS, yielding splittings at least a factor of 2 larger than those arising entirely from barrier alloy disorder in otherwise high symmetry dots. We thus focus on controlled geometrical perturbations of the QD shape to investigate the consequences of reduced symmetry. A similar modelling approach has been successfully employed in the literature~\cite{KaDu2010}.  We target optical selection rules and spectra and these with reported experimental results. This comparison is further supported by group theoretical considerations. Throughout this work, labels such as \sym{C}{3v}, \sym{C}{2v}, and \sym{C}{s} denote the underlying spatial point group symmetries. Since the carrier states are spinorial, their symmetry classification and the corresponding optical selection rules are obtained using the associated double-group representations. The extension to magnetic point groups at finite field is introduced explicitly in Sec.~\ref{sec:ResMag}.

\input{FIGURES/2_sp_states/figure}

\subsection{\label{sec:Modelgeo}Model Geometry and Zero-Field Symmetry}

The site-controlled [111]-oriented GaAs QDs considered theoretically here are based on structures grown and characterised experimentally in Ref.~\cite{RaJu2021a}. The QD confinement geometry is approximated by an equilateral triangular prism with base length $L=26\,\text{nm}$ and height $h = 3\,\text{nm}$. For the ideal structure, the triangular cross section in the (111)-plane is taken to remain constant throughout the dot, giving the confinement geometry \sym{D}{3h} symmetry. The chosen dimensions and alloy compositions are representative of the experimentally relevant size range for these structures~\cite{RaJu2021a, DiPe2012}. The GaAs QD is intersected by a vertical quantum wire (VQWr) of identical cross section composed of Al$_{0.08}$Ga$_{0.92}$As. The QD and VQWR structure is embedded in an Al$_{0.45}$Ga$_{0.55}$As matrix, motivated by experimental findings~\cite{RaJu2021a, DiPe2012}. The lateral quantum wires present in the experimental structure are not included explicitly here. Their inclusion does not alter the symmetry of the structure relevant to the present analysis, although they may modify quantitative features such as, transition energies and $g$-factors. A schematic illustration of the dot system is shown in Fig.~\ref{fig:vpol_111_comparison}~a) with the VQWR extending along the [111]-direction throughout the simulation cell. To introduce a controlled symmetry reducing perturbation, one corner of both the QD and VQWr is truncated as illustrated in Fig.~\ref{fig:vpol_111_comparison}~b). As discussed below, this perturbation leads to a non-vanishing FSS, even in the case of a pure GaAs QD. A similar symmetry reducing perturbation was employed by Karlsson \emph{et al.}~\cite{KaDu2010} to represent deviations from the ideal faceting of the site-controlled structure. Here, the truncation is used in the same spirit as a controlled means of lowering the confinement symmetry. Our theoretical framework will now be applied to these structures to obtain the electronic and excitonic properties in the presence and absence of magnetic fields.

Although the ideal confinement geometry shown in Fig.~\ref{fig:vpol_111_comparison}~a) possesses \sym{D}{3h} symmetry, the exact symmetry of the complete zero magnetic field Hamiltonian is reduced to \sym{C}{3v}. Bulk-inversion asymmetry (BIA) of the zincblende lattice, explicitly included through the $C_k$ terms in the \kp Hamiltonian (see supplementary information), removes the horizontal mirror plane, $\sigma_{xy}$, present when considering the geometric confinement alone. Additionally, growth along the [111] direction gives rise to a built-in piezoelectric potential, \(V_p\) (see Fig.~\ref{fig:vpol_111_comparison}), which also breaks the horizontal mirror symmetry of the confinement geometry. Consequently, \sym{D}{3h} is retained here only as an idealized higher-symmetry reference. As shown in Fig.~\ref{fig:vpol_111_comparison}~c), however, the associated potential drop is much weaker in the GaAs/(Al,Ga)As and In${_{0.15}}$Ga${_{0.85}}$As/GaAs QDs compared to an InAs/GaAs QD of the same size. We note that the reversal of the polarization-field direction in the InAs/GaAs system relative to GaAs/(Al,Ga)As originates from the much stronger second-order piezoelectric contribution associated with the larger lattice mismatch in this QD. Hence, for GaAs/(Al,Ga)As and low-In-content (In,Ga)As/GaAs QDs, the symmetry-breaking effect of the built-in piezoelectric field remains comparatively weak. Optical transitions that are forbidden under the idealized \sym{D}{3h} reference symmetry can become symmetry allowed under the exact \sym{C}{3v}symmetry, but their oscillator strengths need not be appreciable. The calculated optical response may consequently appear \sym{D}{3h}-like when these additional symmetry allowed transitions carry only weak oscillator strength.

Focussing first on the symmetry arising purely from the truncated confinement geometry shown in Fig.~\ref{fig:vpol_111_comparison}~b), and neglecting BIA and polarisation fields, the structure retains a vertical mirror plane, $\sigma_{xz}$, a horizontal mirror plane, $\sigma_{xy}$, and an in-plane \sym{C}{2x} rotation axis. The symmetry imposed solely by the confinement geometry is therefore \sym{C}{2v}. Accounting for BIA and piezoelectric contributions, $\sigma_{xy}$, and consequently also \sym{C}{2x}, are lost, leaving only the vertical mirror plane $\sigma_{xz}$. The exact symmetry of the full zero-field Hamiltonian is thus reduced to \sym{C}{s}. The geometrical truncation therefore does not by itself determine the exact symmetry of the full Hamiltonian, but instead controls the departure of the confinement geometry from the high symmetry trigonal limit, see Fig.~\ref{fig:vpol_111_comparison}. In the [111]-oriented site-controlled GaAs/(Al,Ga)As QDs considered here, the symmetry breaking effects associated with the BIA and piezoelectric terms are comparatively weak, while the truncation provides the dominant tunable perturbation to the confinement geometry. As shown below, varying the truncation factor consequently controls the calculated excitonic FSS within the present model.

\input{FIGURES/3_exciton_config/figure}

In summary, the horizontal mirror plane, $\sigma_{xy}$, is not an exact symmetry of the full Hamiltonian. However, the terms responsible for breaking this symmetry may produce only weak modifications to the calculated optical response for GaAs/(Al,Ga)As and low-In-content (In,Ga)As/GaAs QDs. This motivates the use of \sym{D}{3h} and \sym{C}{2v} as higher-symmetry reference groups for systems whose exact Hamiltonian symmetries are \sym{C}{3v} and \sym{C}{s}, respectively. Importantly, transitions forbidden by a higher symmetry reference group may become symmetry allowed under the exact lower symmetry without necessarily acquiring appreciable oscillator strength. Whether the calculated spectra exhibit approximately \sym{D}{3h}-like or \sym{C}{2v}-like behaviour is therefore assessed below from the oscillator strengths of the symmetry-allowed transitions.

\subsection{Single Particle Electronic Structure}

Before turning to selection rules and optical spectra, we first investigate the bound single particle states of the site-controlled [111]-oriented GaAs/(Al,Ga)As dots discussed in Sec.~\ref{sec:Modelgeo}. For truncations up to $50\%$, we find one bound electron Kramers degenerate pair, \pe, and three bound hole Kramers degenerate pairs, \phz, \pho, \pht, thus two electron and six hole bound states. For larger truncations, the ordering of the optically relevant hole states changes. We therefore restrict the following analysis to the regime before this reordering, since truncations of up to $50\%$ already yield FSS values comparable to those observed experimentally, as discussed later. Figure~\ref{fig:sp_plots} displays the corresponding charge densities, single particle energies and spinor compositions in the [111] $\ket{J,m_j}$ basis for a QD with a truncation of $30\%$. The zero of energy is set to the unstrained GaAs valence band edge.

The hole ground state, \phz, is predominantly HH-like in character (HH $\approx 80\%$). In contrast, the first excited hole state, \pho, exhibits pronounced HH-LH mixing. Its charge density reveals a torus-like envelope-function character. Evaluating the dipole matrix elements following the methodology outlined in Sec.~\ref{sec:Num_Mod}, we find that the transition matrix element between \pho and \pe is close to zero. This behaviour is consistent with observations in systems of similar symmetry, such as III-N QDs, where torus-like hole states have likewise been found to exhibit strongly suppressed optical activity~\cite{BaSc2007}. The state \pho is therefore effectively spectroscopically dark. The second excited hole state, \pht, has a pronounced LH character of approximately $90\%$. The presence of such a strongly LH-dominated bound state close to the valence-band maximum is noteworthy and is uncommon in conventional Stranski-Krastanov QDs~\cite{Zi2013, ScWi2007}.

Before turning to the calculation of excitonic states, we note that the main features of the calculated single-particle electronic structure are consistent with available experimental observations. Spectroscopic studies of site-controlled [111]-oriented (In,Ga)As/(Al,Ga)As QDs, modelled in Ref.~\cite{KaDu2010} using a uniform In concentration of $10\%$, indicate the presence of a single bound electron state and two optically relevant bound hole states, one predominantly HH-like and the other predominantly LH-like in character. In our calculations, the additional torus-like hole state, \pho, should basically be spectroscopically dark due to its strongly suppressed dipole matrix element with the electron ground state. Consequently, despite the presence of three bound hole states in the single-particle spectrum, only two are expected to contribute directly to the experimentally accessible optical transitions and the associated excitonic complexes. Thus in simplified single-particle picture, the calculated electronic structure is in qualitative agreement with the experiment. In the following section, we account for Coulomb effects and evaluate excitonic complexes in detail.

\subsection{Zero-Magnetic Field: Excitonic States and Selection Rules}

To classify the excitonic states, we adopt the notation of Ref.~\cite{KaDu2010}, \sym{X}{\alpha\beta}, where $\alpha$ and $\beta$ denote the occupations of the (two optically active) bound hole states. Since only a single bound electron state is present, the notation implicitly assumes occupation of the electron ground state. We stress that this labelling provides only a simplified classification of the excitonic states. Within the CI approach, the many-body eigenstates are, in general, linear combinations of Slater determinants and therefore contain contributions from several single-particle hole configurations. Nevertheless, the notation has a practical use in providing a convenient means of identifying the dominant configuration and facilitates a direct comparison with experimental literature~\cite{KaDu2010}.

In the following, we focus on the two lowest-energy bright excitonic complexes, \sym{X}{10} and \sym{X}{01}. The ground-state exciton, \sym{X}{10}, is dominated by the configuration involving the electron ground state, \pe, and the HH-dominant hole state, \phz, whereas the excited exciton, \sym{X}{01}, is dominated by the configuration formed from \pe and the LH-dominant hole state, \pht. A schematic illustration of this excitonic-state labelling is given in Fig.~\ref{fig:geom_config}. As discussed above, the intermediate hole state, \pho, is omitted from the labelling since our CI calculations confirm that the associated excitonic complex is effectively spectroscopically dark. The subscripts therefore refer to the occupations of the two optically active bound states, \phz, \pht.

\input{FIGURES/4_exciton_rules/figure}

In what follows, we assign the labels \sym{X}{10} and \sym{X}{01} only when the corresponding single-particle state configuration accounts for more than $95\%$ of the CI state. The predominantly HH- and LH-like character of the relevant hole states motivates a group-theoretical analysis of the optical selection rules expected for HH- and LH-excitonic states under different point-group symmetries. Figure~\ref{fig:X_B_0} summarises these selection rules for the \sym{D}{3h}, \sym{C}{3v}, \sym{C}{2v}, and \sym{C}{s} point groups. We note that this analysis assumes pure HH and LH states, such that HH-LH mixing is neglected. The role and importance of band-mixing effects in the underlying single-particle states are examined in more detail below.

In an idealised picture, the neutral exciton is formed from a conduction-band electron, $\left(\ket{1/2,\pm1/2}_c = |\text{CB}\pm\rangle\right)$, and a valence-band hole state. For a HH exciton, the relevant hole basis states are $\ket{3/2,\pm3/2}_v$. In systems with \sym{D}{3h} or \sym{C}{3v} symmetry, the four resulting excitonic states form two doublets, as illustrated in Fig.~\ref{fig:X_B_0}~a). For \sym{D}{3h} symmetry, one of these doublets is optically active through the in-plane dipole components ($x,y$), whereas the other remains dark. For light propagating along the growth direction, $z$, these transitions therefore correspond to light polarized within the $x$--$y$ plane. Under \sym{C}{3v} symmetry, both doublets possess symmetry allowed optical transitions, coupling to the in-plane ($x,y$) optical polarization; the transitions allowed in \sym{C}{3v} but forbidden under \sym{D}{3h} symmetry are indicated by dashed lines. In either case, the doublet degeneracy precludes a splitting between the $x$- and $y$-polarised bright excitonic states, and therefore the corresponding excitonic FSS vanishes.

When the symmetry is reduced to \sym{C}{2v} or \sym{C}{s}, the doublets split into four singlets, as indicated in Fig.~\ref{fig:X_B_0}~b). For \sym{C}{2v} symmetry, group theory predicts one symmetry forbidden state and three optically active states, with one bright state in $x$-, one in $y$-, and one in $z$-polarisation. Conventionally, the splitting between the two in-plane $x$- and $y$-polarized bright states defines the FSS for a pure HH exciton, $\Delta_{\text{fss}}^{\text{HH}}$. Reducing the symmetry further to \sym{C}{s} permits additional optical channels that are forbidden under \sym{C}{2v}, as indicated by the dashed lines in Fig.~\ref{fig:X_B_0}~b). For the truncated QD structure considered here, Fig.~\ref{fig:vpol_111_comparison}~b), the remaining mirror plane is $\sigma_{xz}$, which pins the polarization of the additional symmetry allowed transitions to be either within the mirror plane ($x,z$) or perpendicular to it ($y$). The equivalent result for a $\sigma_{yz}$ mirror plane is obtained by interchanging the $x$ and $y$ assignments~\cite{KaDu2010}. In principle, given the selection rules, different FSS splittings for $x$-$y$ polarised transitions can be defined in the \sym{C}{s} symmetric case. For illustration purposes we have used the pair of states as in the \sym{C}{2v}-symmetric case.

The LH exciton, formed from a conduction-band electron and the hole basis states $\ket{\frac{3}{2},\pm\frac{1}{2}}_v$, exhibits the selection rules shown in Fig.~\ref{fig:X_B_0}~c), d). For both \sym{D}{3h} and \sym{C}{3v} symmetry, the four LH excitonic states comprise one symmetry forbidden singlet, one doublet optically active in the $x$- and $y$-polarised dipole components, and one singlet active through the $z$-polarised dipole component. When the symmetry is lowered to \sym{C}{2v}, the in-plane bright doublet splits into two non-degenerate singlets, giving rise to a corresponding LH FSS, $\Delta_{\text{fss}}^{\text{LH}}$. A further reduction to \sym{C}{s} opens additional optical transition channels, indicated by dashed lines again, that are symmetry forbidden in the \sym{C}{2v} case.

The key distinction for the analysis below is therefore that a finite bright $x$-$y$ excitonic FSS observed in experiment or theory is incompatible with the \sym{D}{3h} and \sym{C}{3v} symmetry groups considered here, whereas it is symmetry allowed for \sym{C}{2v} and \sym{C}{s}. Group theory determines which optical transitions are symmetry allowed or forbidden, but not their absolute oscillator strengths. We therefore use the FSS and symmetry allowed transition pattern to identify compatible exact symmetries, while the calculated oscillator strengths are used to assess whether the dominant optical response resembles that of a higher-symmetry group.

\subsection{Zero-Magnetic Field: Calculated Optical Spectra}

\input{FIGURES/5_numeric_spectra/figure}

Equipped with insight from group theory, we now present the results of our excitonic calculations. We start with the ideal high symmetry structure displayed in Fig.~\ref{fig:vpol_111_comparison}~a) and the obtained \sym{X}{10} and \sym{X}{01} spectra are shown in Fig.~\ref{fig:fss_spec}~a). The HH-dominant \sym{X}{10} spectrum consists of two optically allowed doublets with vanishing $x$-$y$ FSS. The lower energy doublet, however, carries only a very small oscillator strength and is visible only in the inset of Fig.~\ref{fig:fss_spec}~a) after magnification by a factor of $10^6$. These transitions are symmetry allowed under \sym{C}{3v} but forbidden in the elevated \sym{D}{3h} point group, see Fig.~\ref{fig:X_B_0}~a). The strong suppression of the oscillator strength of one of the two doublets produces a \sym{D}{3h}-like optical response despite the exact \sym{C}{3v} symmetry of the zero magnetic field Hamiltonian. Throughout this work, we reserve the term ``dark'' for symmetry-forbidden transitions. Numerically, we classify a transition as dark when its calculated oscillator strength is below $10^{-20}$, which we treat as zero within the numerical accuracy of our calculations. The weak, lower energy doublet discussed above is therefore not classified as dark. Although its oscillator strength is strongly suppressed and the corresponding emission may consequently be difficult to resolve experimentally, it remains many orders of magnitude larger than that of a symmetry-forbidden transition. For the LH-dominant \sym{X}{01} complex, the calculated spectrum contains a dark exciton state, an in-plane polarized doublet and a $z$-polarised singlet, see Fig~\ref{fig:fss_spec}~a), again closely reproducing the corresponding \sym{D}{3h}/\sym{C}{3v} pattern of Fig.~\ref{fig:X_B_0}~c).

We next consider the excitonic spectrum of a QD with $30\%$ geometrical truncation, see Fig.~\ref{fig:vpol_111_comparison}~b), for which the corresponding single-particle states were discussed in Sec.~\ref{subsec:SP}. For the HH-dominant \sym{X}{10} complex, the in-plane bright-state degeneracy is lifted, resulting in a finite FSS, $\Delta_{\text{fss}}^{\text{HH}}$. The dominant optical response consists of an $x$- and a $y$-polarised transition, together with a much weaker $z$-polarised transition at lower energy, which becomes visible in the inset after magnification by a factor of $10^3$. These dominant transitions coincide with those allowed by the \sym{C}{2v} reference symmetry shown in Fig.~\ref{fig:X_B_0}~b), whereas the additional channels permitted by the exact \sym{C}{s} symmetry carry strongly suppressed oscillator strengths. Thus, in the absence of a magnetic field, the calculated \sym{X}{10} spectrum exhibits a \sym{C}{2v}-like optical response even though the exact zero-field symmetry of the Hamiltonian is \sym{C}{s}. The LH-dominant \sym{X}{01} complex follows the same pattern, meaning that the observed spectrum may be described by a \sym{C}{2v} symmetric system, following the selection rules summarized in Fig.~\ref{fig:X_B_0}~d) with the additional \sym{C}{s} allowed decay pathways remaining strongly suppressed. The in-plane doublet is split, giving a finite FSS, $\Delta_{\text{fss}}^{\text{LH}}$, while the $z$-polarised singlet remains on the high energy side of the spectrum.

\input{FIGURES/6_hh_lh_sp_states/figure}

Overall, when introducing the dot truncation, our theoretical studies reflect not only the group theoretically expected behaviour, but also experimentally observed optical spectra for both \sym{X}{10} and \sym{X}{01} excitons~\cite{KaDu2010, RaJu2021a}. The calculated FSS value of \sym{X}{10} and \sym{X}{01} depends on the truncation factor, which serves as a tunable parameter of the strength of the symmetry reducing perturbation. Other perturbations, such as dot elongation~\cite{PiBa2023}, that produce the same reduction of the confinement symmetry are therefore expected to yield qualitatively similar optical behaviour. Previous measurements of site-controlled [111]-oriented GaAs QDs have reported \sym{X}{10} FSS values on the $\mu$eV scale~\cite{RaJu2021a}. As shown in Fig.~\ref{fig:fss_spec}~c), the calculated $x$-$y$ FSS remains on this experimentally relevant $\mu$eV scale over a broad range of truncations. In what follows, we adopt a $30\%$ truncation as a representative reduced symmetry structure, resulting in FSS values of $\Delta_{\text{fss}}^{\text{HH}} = 3.84\,\mu\text{eV}$ and $\Delta_{\text{fss}}^{\text{LH}} = 4.53\,\mu\text{eV}$.

Before turning to the magnetic-field response of the GaAs QDs discussed, it is instructive to examine why the numerical results, which fully account for band-mixing effects, nevertheless closely follow the group-theoretical predictions derived under the assumption of pure HH and LH states. To shed light on this point, we briefly analyse the band decomposition of the relevant single-particle states and discuss the resulting consequences for the excitonic states in the following section.

\subsection{Band Mixing and Optical Selection Rules}

The close agreement between the calculated spectra and the group theoretically predicted reference pattern using idealised HH- and LH-exciton states may initially be surprising, since as already mentioned in the previous section, the underlying single particle hole states exhibit appreciable HH-LH mixing. As shown in Fig.~\ref{fig:sp_plots}, \phz has approximately $80\%$ HH character, while \pht exhibits a LH character of approximately $90\%$. To determine how band mixing contributes to the optical response, we resolve \phz and \pht into their HH and LH envelope function components and examine their spatial distributions in Fig.~\ref{fig:HH-LH}. Contributions from the spin SO bands are omitted from the spatial analysis because their weights in the calculated spinors are negligible, see Fig.~\ref{fig:sp_plots}.

For the HH-dominant hole ground state, \phz, the corresponding charge-density decomposition, shown in Fig.~\ref{fig:HH-LH}~a), reveals that the majority HH component is strongly localized near the centre of the GaAs QD, whereas the minority LH component is displaced towards the dot periphery and, in particular, towards its corners. Since the bound electron state, \pe, is likewise strongly localized in the central region, see Fig.~\ref{fig:sp_plots}, its spatial overlap with the HH component is substantially larger than with the minority LH component. Consequently, the HH contribution to the hole spinor dominates the corresponding dipole matrix element, and the optical response associated with \pe and \phz is therefore governed primarily by the CB-HH wave function overlap component. This highlights that the spatial distribution of the individual band components, rather than their integrated weights alone, determine the optical response.

To quantify this effect, we define a normalised band resolved optical weight,
\begin{equation}
f(\xi, j) = \frac{W(\xi,j)}{\sum_{\eta}W(\eta,j)}\,,\label{eq:BROW}
\end{equation} where $\xi \in \{\text{CB},\,\text{HH},\,\text{LH},\,\text{SO}\}$ denotes a basis state (band component) and,
\begin{align}
    W(\xi,j) =& \frac{1}{4}\sum_{a\in\mathcal{K}^e_{0}} \sum_{b\in\mathcal{K}^h_{j}} \sum_{i=x,y,z} \nonumber\\&\times \left|\left\langle\psi_{e,a}\left|\left.\frac{\partial H_{\mathbf{k\cdot p}}}{\partial k_i}\right|_{\mathbf{k}=\mathbf{0}}\right|\psi_{h,b}^\xi\right\rangle\right|^2\,.
    \label{eq:BandOverlapII}
\end{align}

Here $\mathcal{K}^{e}_{0}$ and $\mathcal{K}^{h}_{0}$ denote the electron and hole Kramers doublets, respectively, and $\psi_{h,b}^{\xi}$ is the component of the 8-band hole spinor belonging to band $\xi$. The factor of $1/4$ averages over the four combinations of electron and hole Kramers partners removing the arbitrary choice of basis within each degenerate subspace, while the sum over $x$, $y$ and $z$ accounts for optical selection rules. We evaluate $f(\xi,j)$ for the manifold of bright hole states, $j\in\{0,2\}$.

\input{Table/GT_Groups}

For \phz, the resulting optical weights are \mbox{$f(\mathrm{HH},0)=0.995$} and \mbox{$f(\mathrm{LH},0)=0.005$}, demonstrating that the dipole matrix element and as such the optical selection rules are almost entirely HH-dominant despite the appreciable LH contribution to the underlying single-particle wave function. The opposite behaviour is found for the LH-dominant state, \pht, shown in Fig.~\ref{fig:HH-LH}~b). Its majority LH component is concentrated in the central region of the GaAs QD, whereas the minority HH contribution is displaced towards the dot corners. Consequently, $f(\mathrm{LH},2)=0.999$, while $f(\mathrm{HH},2)=0.001$, highlighting that for \pht the optical selection rules are almost entirely determined by the LH character of \pht. These strongly HH- and LH-dominant optical weights persist throughout the range of dot truncations considered and are therefore not specific to the $30\%$ truncation value considered. 

From the above, we conclude that, despite appreciable HH-LH mixing at the single-particle level, the optical selection rule patterns can remain well described in terms of pure HH and LH states. All this explains why the excitonic complexes \sym{X}{10} and \sym{X}{01} obtained from our CI calculations can, to a very good approximation, be described as predominantly pure HH- and LH-like complexes, respectively; thus they follow closely the corresponding idealised group theoretical selection rules. This observation provides a basis for analyzing the magneto-optical response of site-controlled [111]-oriented GaAs QDs in the following section.

%% file: FIGURES/2_sp_states/figure.tex
\begin{figure}[t]
    \includegraphics[width=\columnwidth]{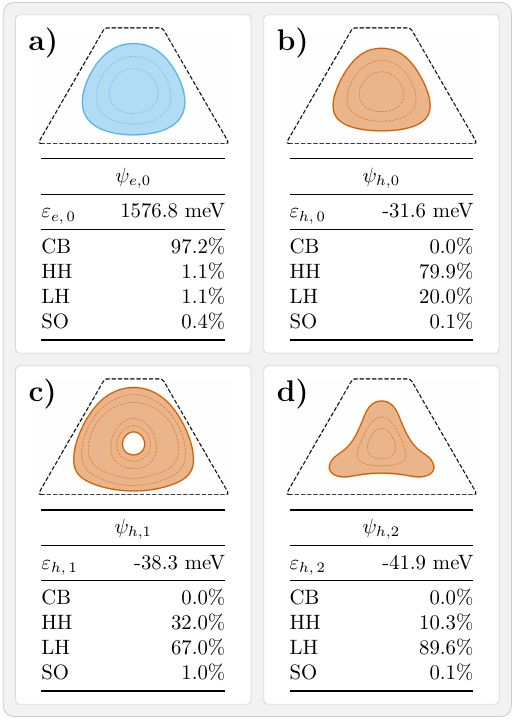}
    \caption{Charge densities and energies of the bound single particle states obtained for a [111]-oriented site-controlled GaAs/(Al,Ga)As prismatic quantum dot including the symmetry lowering effects of a corner truncation. The table also provides spinor decompositions in the total angular momentum basis classified by conduction (CB), heavy-hole (HH), light-hole (LH) and split-off (SO) band components. a) The electron ground state, \pe. b) The HH dominant hole ground state, \phz. c) The torus-like first excited hole state, \pho. d) The LH dominant second excited hole state, \pht.}
    \label{fig:sp_plots}
\end{figure}

%% file: FIGURES/3_exciton_config/figure.tex
\begin{figure}[t]
    \includegraphics[width=\columnwidth]{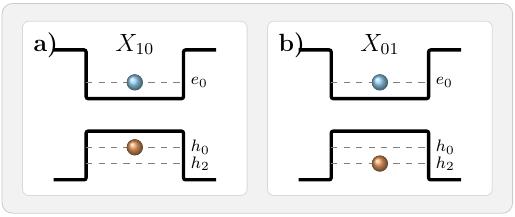}
    \caption{Schematic illustration of dominant single particle configurations of the energetically lowest bright excitonic complexes denoted (a) \sym{X}{10} and (b) \sym{X}{01}. The label $e_0$ refers to the electron ground state, while $h_0$ and $h_2$ denote the heavy-hole (HH) ground and light-hole (LH) like excited hole states.}
    \label{fig:geom_config}
\end{figure}

%% file: FIGURES/4_exciton_rules/figure.tex
\begin{figure}[t]
    \includegraphics[width=\columnwidth]{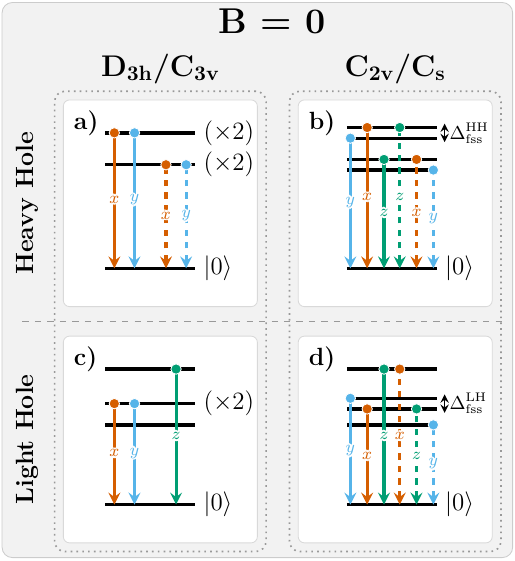}
     \caption{Diagramatic illustration of optical selection rules for a neutral exciton at zero magnetic field ($\mathbf{B} = \mathbf{0}$) for systems with different point group symmetries. a) and c) shows allowed excitonic transitions for heavy-hole (HH) and light-hole (LH) excitons respectively: extra transitions allowed under \sym{C}{3v} but forbidden under \sym{D}{3h} symmetry are shown as dashed lines. b) and d) shows allowed excitonic transitions for HH and LH excitons: extra transitions allowed under \sym{C}{s} but forbidden under \sym{C}{2v} symmetry are shown as dashed lines. $x$, $y$ and $z$ denote the light polarization characteristics of emitted photons in a recombination process.}
    \label{fig:X_B_0}
\end{figure}

%% file: FIGURES/5_numeric_spectra/figure.tex
\begin{figure}[tp]
    \includegraphics[width=\columnwidth]{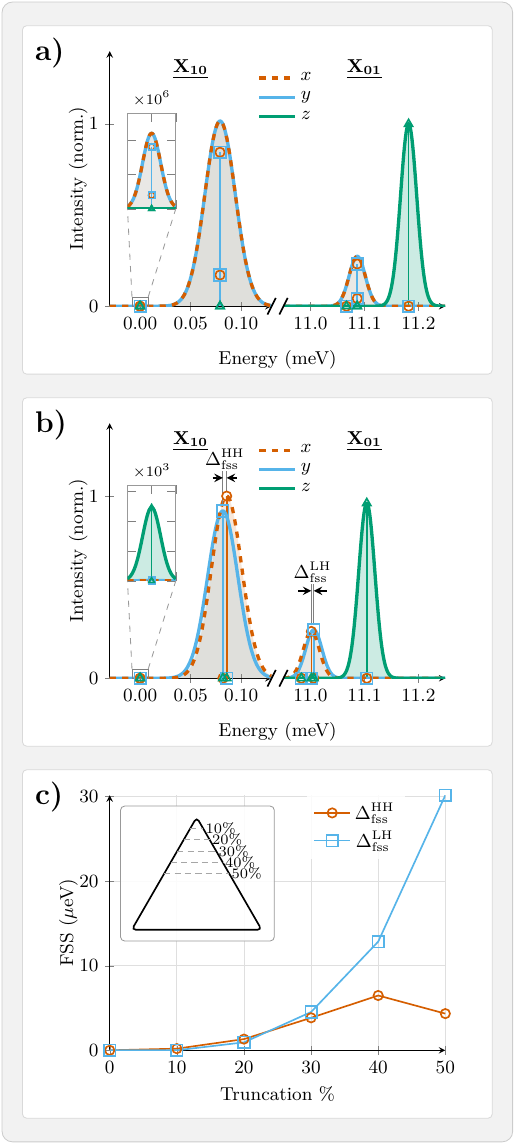}
    \caption{Calculated excitonic spectra of GaAs QDs. In a) and b), stems show $x$-, $y$- and $z$-polarised transition strengths, curves show Gaussian broadened profiles ($\sigma=15\,\mu\mathrm{eV}$). Energies relative to lowest energy \sym{X}{10} transition ($1584.6$ and $1586.0$~meV). Intensities are normalised to the brightest line in each panel. a) Ideal \sym{C}{3v} dot with vanishing FSS. Inset: weak $x$-$y$ doublet ($\times10^{-6}$). b) Dot with $30\%$ truncation: both \sym{X}{10} and \sym{X}{01} split into one dark and three bright ($x,y,z$) singlets. Inset: weak $z$-polarised \sym{X}{10} line ($\times10^{-3}$). c) $x$-$y$ FSS of \sym{X}{10} and \sym{X}{01} versus corner truncation. Inset: Truncation planes.}
    \label{fig:fss_spec}
\end{figure}

%% file: FIGURES/6_hh_lh_sp_states/figure.tex
\begin{figure}[t]
    \includegraphics[width=\columnwidth]{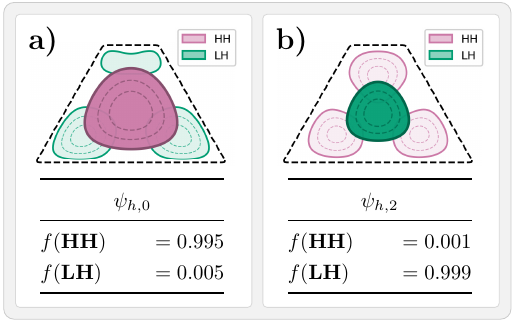}
    \caption{Spatial distribution of heavy hole (HH) and light-hole (LH) components and their related optical weights, Eqs.~\eqref{eq:BROW} and \eqref{eq:BandOverlapII}, for the two bright bound hole states a) \phz (HH dominant) and b) \pht (LH dominant) for a site-controlled [111]-oriented GaAs QD with $30\%$ truncation.} 
    \label{fig:HH-LH}
\end{figure}

%% file: Table/GT_Groups.tex
\setlength\cellspacetoplimit{5pt}
\setlength\cellspacebottomlimit{5pt}
\newcommand{\cellstack}[1]{\begin{tabular}[c]{@{}c@{}}#1\end{tabular}}

\begin{table}[t]
\setlength{\tabcolsep}{4pt}
\begin{tabular}{|Sc|Sc|Sc|}
\hline
\cellstack{$G$\\Parent\\group} & \cellstack{$H$\\Unitary\\operations} & \cellstack{$\mathcal{T}(G-H)$\\Anti-unitary\\operations}\\ \hline
\hline
\cellstack{\sym{C}{3v}\\$=\{E, C_3, C_3^2, 3\sigma_v\}$} & \cellstack{\sym{C}{3}\\$=\{E, C_3, C_3^2\}$} & $\mathcal{T}(\sigma_v)$\\ \hline
\cellstack{\sym{C}{s}\\$=\{E, \sigma_{xz}\}$} & \cellstack{\sym{C}{1}\\$=\{E\}$} & $\mathcal{T}(\sigma_{xz})$\\ \hline
\cellstack{\sym{C}{2v}\\$=\{E, C_{2x}, \sigma_{xy}, \sigma_{xz}\}$} & \cellstack{\sym{C}{s}\\$=\{E, \sigma_{xy}\}$} & $\mathcal{T}(\sigma_{xz}, C_{2x})$\\ \hline
\end{tabular}
\caption{Magnetic point groups relevant to the present [111]-oriented geometry for an applied field $\mathbf{B}\parallel[111]\parallel \mathbf{\hat{z}}$. The unitary subgroup, $H$, contains the spatial operations that leave the axial field $B_z$ invariant. Operations in $G-H$ reverse $B_z$ and only remain symmetries when combined with the time-reversal operator, $\mathcal{T}$. In the \sym{C}{3v} row, $\sigma_v$ denotes the three vertical mirror operations allowed in the group.}
\label{tab:sumaryGroupMagnet}
\end{table}

%% file: sections/3_Results/magnetic.tex
\section{\label{sec:ResMag} Faraday Geometry Optical Properties}

Having established that the exact \sym{C}{s} structure exhibits a dominant \sym{C}{2v}-like optical response at zero field, we now examine how this description evolves when a magnetic field is applied along the [111] growth direction, i.e., in Faraday geometry. We first determine the magnetic group selection rules associated with \sym{C}{3v} and \sym{C}{s} groups as these are the exact symmetries of the dots studied here. Furthermore, given that literature results often use the \sym{C}{2v} symmetry to explain optical spectra in the absence of magnetic fields, we also study this point group. This analysis allows us to test whether the \sym{C}{2v}-like description identified at zero-field can also account for the characteristic four transition trion response found experimentally in Faraday geometry. We then compare this reference description with the exact magnetic symmetry of the truncated, reduced symmetry QD.

\subsection{Faraday Geometry: General Selection Rules}

To interpret the trion spectrum at finite magnetic field, we first summarise the magnetic point groups and selection rules, which will be used as a reference for our numerical simulation. For the type III magnetic point groups relevant here, we use the notation~\cite{BrCr2010}
\begin{equation}
    G(H)= H + \mathcal{T}(G-H)\,,
    \label{eq:TypeIIIMag}
\end{equation}
where $G$ is the parent geometric point group in the absence of a magnetic field, $H$ is the subgroup whose operations remain unitary at fixed magnetic field, and $\mathcal{T}$ is the time reversal operator. Operations in $G-H$ reverse the applied magnetic field and therefore remain symmetries only when combined with $\mathcal{T}$. Since the magnetic field is an axial vector, its transformation under the spatial symmetry operations determines the appropriate unitary subgroup~\cite{BrCr2010}.

For the present [111]-oriented \sym{C}{3v} symmetric reference system and when assuming Faraday geometry, the magnetic field, $\mathbf{B}\parallel[111]\parallel\mathbf{z}$, is aligned with the threefold rotation axis, \sym{C}{3}. The \sym{C}{3} rotations leave the axial field, $B_z$ invariant, whereas the vertical mirror operations reverse it and remain symmetries only when combined with time reversal operation, see Table~\ref{tab:sumaryGroupMagnet}. The corresponding magnetic point group is therefore \sym{C}{3v}{}(\sym{C}{3}). For a positively charged trion, $X^+$, composed of one electron and two HH valence states, this symmetry permits the relevant mixing of the $\ket{3/2,+3/2}$ and $\ket{3/2, -3/2}$ HH components~\cite{DuGl2013}. The resulting trion spectrum contains four optically allowed transitions, $\nu_1,\dots,\nu_4$, with two transitions of each circular polarisation, as summarised in Fig.~\ref{fig:Xp_B_selection_rules}~a).

For comparison, consider the conventional \sym{C}{2v}{}(\sym{C}{2}) case in which the two-fold rotation axis is parallel to the applied magnetic field, as for [001]-oriented QDs in Faraday geometry with $\mathbf{B}\parallel[001]$. This magnetic symmetry forbids mixing between the predominantly HH states and consequently permits only two optically active trion transitions, one in $\sigma^+$ and one in $\sigma^-$ polarisation~\cite{BaGe2011}. This conventional \sym{C}{2v}{}(\sym{C}{2}) description does not necessarily apply to the \sym{C}{2v} reference symmetry relevant to the present [111]-oriented QDs, for which the \sym{C}{2} rotation axis lies in-plane and perpendicular to the applied field. The corresponding higher-symmetry reference at finite field is therefore \sym{C}{2v}{}(\sym{C}{s}), which we consider next.

\input{FIGURES/7_trion_B_selection_rules/figure}

For the \sym{C}{2v} reference symmetry used to describe the dominant zero magnetic field optical response, the relevant symmetry operations are $\mathrm{C}_{\mathrm{2v}} = \{E, \sigma_{xy}, \sigma_{xz}, \mathrm{C}_{2x}\}$ for our [111]-oriented dots with $\mathbf{B}\parallel[111]\parallel \mathbf{z}$. Here, the \sym{C}{2x} rotation axis lies in the QD plane and is therefore perpendicular to the Faraday geometry magnetic field. At fixed $\mathbf{B}$, the operations $E$ and $\sigma_{xy}$ leave the axial magnetic field invariant, whereas \sym{C}{2x} and $\sigma_{xz}$ reverse it and are restored only by time reversal symmetry. The corresponding magnetic point group is therefore, \sym{C}{2v}(\sym{C}{s}), with unitary subgroup, \mbox{\sym{C}{s}$=\{E, \sigma_{xy}\}$}, see Table~\ref{tab:sumaryGroupMagnet}.

The resulting $X^+$ selection rules are shown in Fig.~\ref{fig:Xp_B_selection_rules}~c), with their full derivation given in the supplementary information. Of the four possible transitions, only two are allowed with in-plane circular polarisation, $\sigma^\pm$, while the remaining two are allowed only in $z$ polarisation. The \sym{C}{2v}{}(\sym{C}{s}) higher-symmetry reference thus predicts two in-plane polarised transitions. This provides a distinctly different selection rule pattern compared to \sym{C}{3v}{}(\sym{C}{3}).

Finally, we consider the finite magnetic field symmetry obtained from the exact zero-field \sym{C}{s} Hamiltonian of the truncated QD. At $\mathbf{B}=\mathbf{0}$, the only non-trivial spatial symmetry operation is the vertical mirror plane, $\sigma_{xz}$. Under $\mathbf{B}\parallel [111] \parallel\mathbf{z}$, this reflection reverses the axial magnetic field and therefore only remains as an anti-unitary symmetry. The identity $E$ is the only unitary symmetry operation, giving the magnetic point group \sym{C}{s}{}(\sym{C}{1}), with \mbox{\sym{C}{1} $= \{E\}$}.

The resulting $X^+$ selection rules are shown in Fig~\ref{fig:Xp_B_selection_rules}~b), with their full derivation given in the supplementary information. In contrast to the \sym{C}{2v}{}(\sym{C}{s}) case, \sym{C}{s}{}(\sym{C}{1}) symmetry permits all four HH trion transitions with in-plane optical activity, while $z$-polarised components are also symmetry allowed. The surviving anti-unitary symmetry, $\mathcal{T}(\sigma_{xz})$, constrains the $x$- and $y$-dipole amplitudes to be exactly $90^\circ$ out of phase, such that the in-plane optical transitions are, in general, elliptically polarised.

The above group-theoretical analysis provides the general symmetry framework for the magneto-optical response of the different point groups. However, as before, symmetry alone does not determine the corresponding oscillator strengths and therefore cannot by itself resolve the apparent discrepancy in the literature in terms of the symmetry groups used to explain optical properties of [111]-oriented QDs in the absence and presence of magnetic fields. The numerical calculations presented below are required to establish whether a consistent and unified description of the optical properties of [111]-oriented GaAs QDs can be obtained, thereby reconciling the apparently contradictory observations of \sym{C}{2v}-like behaviour at zero magnetic field and \sym{C}{3v}{}(\sym{C}{3})-like behaviour in Faraday geometry.

The \sym{C}{s}{}(\sym{C}{1}) symmetry therefore permits all four trion transitions in in-plane polarisation in Faraday geometry. Together with the zero-field analysis, this establishes the symmetry allowed optical response of the same reduced symmetry QD in the presence and absence of a magnetic field.

\subsection{Calculated Faraday Geometry Optical Response}

We now calculate the magneto-optical response of the representative $30\%$ truncated dot structure established in Sec.~\ref{sec:Res1}. As discussed above, the system has \sym{C}{s} symmetry, but the zero magnetic field optical properties can be well approximated by the higher symmetry reference \sym{C}{2v}. The single particle and many-body states are here re-evaluated for magnetic fields $0\leq B\leq9\,\text{T}$ applied along the [111] growth direction. To maintain a consistent basis across the zero- and finite-field calculations, we retain the same bound state CI basis comprising two electron and six hole states, with the self-consistent Hartree procedure of Sec.~\ref{sec:Num_Mod}.

\input{FIGURES/8_trion_numerics/figure}

The calculated polarisation resolved ground trion spectrum is shown in Fig.~\ref{fig:bfield}~a). Indeed, four dominant optical transitions are obtained, whose helicity assignments closely follow the \sym{C}{3v}{}(\sym{C}{3}) reference pattern. The inset, however, reveals that each transition also carries a weak component of the opposite circular polarisation together with a separate $z$-polarised contribution, as permitted by the exact \sym{C}{s}{}(\sym{C}{1}) magnetic symmetry. The opposite helicity components are suppressed to approximately the $10^{-3}$ level relative to their corresponding dominant circular components. Consequently, although the exact magnetic group is \sym{C}{s}{}(\sym{C}{1}), the dominant oscillator strengths organise into a \sym{C}{3v}{}(\sym{C}{3})-like selection rule pattern. Furthermore, the energetic ordering of the dominant $\sigma^+$- and $\sigma^-$-polarised transitions is also consistent with that reported experimentally in Ref.~\cite{BaSa2025a}.

Figure~\ref{fig:bfield}~b) shows the evolution of the four transitions, $\nu_1,\,\nu_2,\,\nu_3,\,\nu_4$, across the modelled finite-field range. Their dominant helicity assignments remain unchanged throughout, while the minority opposite helicity components remain strongly suppressed. The calculated finite-field optical response is therefore incompatible with the \sym{C}{2v}{}(\sym{C}{s}) reference, which permits only two in-plane polarised HH trion transitions. Thus, although the zero-field optical response of the truncated QD is well approximated by the higher symmetry \sym{C}{2v} reference, this description breaks down in Faraday geometry. The exact \sym{C}{s}{}(\sym{C}{1}) magnetic symmetry instead provides a consistent symmetry description, while the dominant calculated oscillator strengths exhibit a \sym{C}{3v}{}(\sym{C}{3})-like pattern, thus in agreement with experimental studies.

The four dominant trion transitions obtained numerically are strongly near circular, despite the more general elliptically polarised in-plane response permitted by the exact magnetic symmetry of the system. The extra weak opposite helicity components are permitted but the less restrictive constraints of the \sym{C}{s}{}(\sym{C}{1}) magnetic group. In the present calculations these transitions remain close to the circular polarisation limit and exhibit only very weak ellipticity. A detailed analysis of these transitions is provided in the supplementary information.

The apparent contrast between the zero and finite magnetic field regimes reported in the literature and reproduced here therefore does not require any restoration or elevation of the structural symmetry. The same truncated QD is described by an exact \sym{C}{s} symmetry at zero magnetic field and exact \sym{C}{s}{}(\sym{C}{1}) magnetic symmetry in Faraday geometry, while its dominant optical response changes from \sym{C}{2v}-like to \sym{C}{3v}{}(\sym{C}{3})-like. This also shows that the \sym{C}{2v}-like description appropriate to the zero field regime cannot simply be extended to finite field. A single reduced symmetry structural model can therefore account consistently for the optical behaviour in both regimes.

%% file: FIGURES/7_trion_B_selection_rules/figure.tex
\begin{figure}[t]
    \includegraphics[width=\columnwidth]{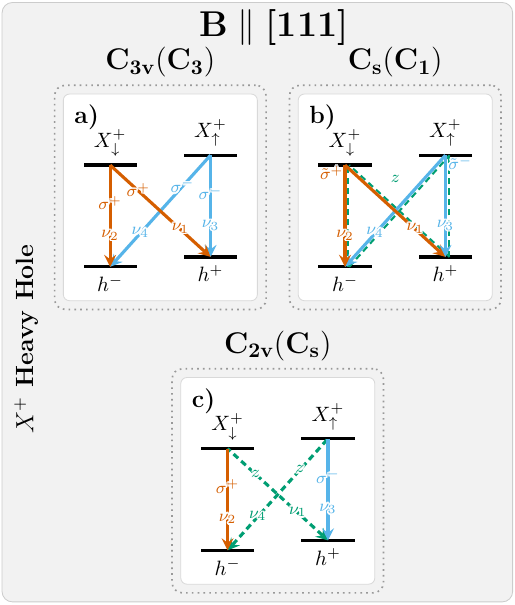}
    \caption{Symmetry derived optical selection rules for the positively charged trion ($X^+$) under a magnetic field, $\mathbf{B} \parallel [111]$, in Faraday geometry, shown for the a) \sym{C}{3v}{}(\sym{C}{3}), b) \sym{C}{s}{}(\sym{C}{1}) and c) \sym{C}{2v}{}(\sym{C}{s}) point group symmetries. The four possible transitions are labelled $\nu_1$--$\nu_4$. Dashed green lines denote $z$-polarised emission, $\sigma^{\pm}$ denotes circular polarization, while elliptical polarisation is labelled as $\tilde\sigma^{\pm}$; the dominant helicity is indicated via the $\pm$ superscript.}
    \label{fig:Xp_B_selection_rules}
\end{figure}

%% file: FIGURES/8_trion_numerics/figure.tex
\begin{figure}[b]
    \includegraphics[width=\columnwidth]{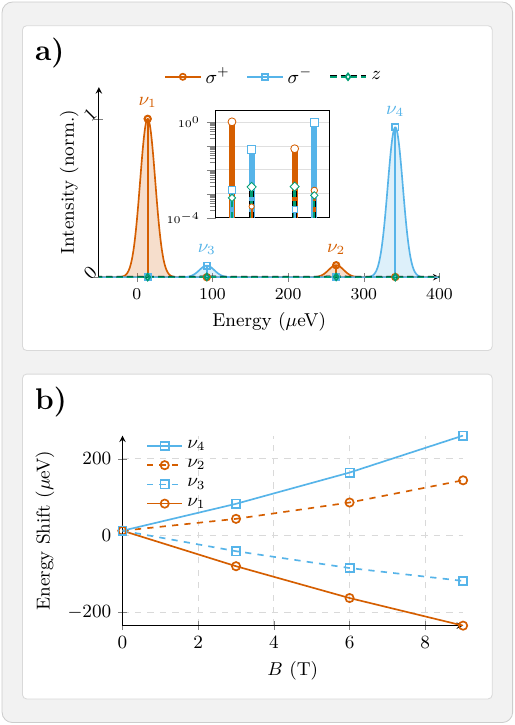}
    \caption{Calculated magneto-optical response of the $X^+$ in the $30\%$ truncated GaAs QD. a) Spectrum at $6\,\mathrm{T}$. Stems show $\sigma^\pm$- and $z$-polarised transition strength, curves show Gaussian broadened profiles ($\sigma=10\,\mu\mathrm{eV}$). Energies relative to the zero field $\nu_1$ transition ($1589.3\,\mathrm{meV}$). Intensities normalised to the brightest line. Inset: log scale stems resolving weak opposite helicity and $z$ components. b) Energies of $\nu_1$-$\nu_4$ versus $B$, relative to their mean at each field to remove the diamagnetic shift~\cite{BaSa2025a}. Note the different energy zero from a).}
    \label{fig:bfield}
\end{figure} 

%% file: sections/4_Conclusion/conclusion.tex
\section{\label{sec:Conc}Conclusion}

This work addresses the apparently conflicting symmetry signatures reported for the zero and finite magnetic field optical response of [111]-oriented site-controlled arsenide-based QDs. By combining group theoretical analysis with 8-band \kp theory and a self-consistent Hartree informed CI calculations of the many-body states, we show that both regimes can be described consistently using the same reduced symmetry structure.

At zero magnetic field, introducing a controlled symmetry reducing perturbation, similar to previous works, the exact symmetry of the Hamiltonian is lowered to \sym{C}{s}. The resulting optical spectrum nevertheless remains predominantly \sym{C}{2v}-like. We attribute this effect to the weak polarisation field in [111]-oriented GaAs QDs. Overall, our calculations reproduce the experimentally reported zero-field behaviour without requiring the QD to possess exact \sym{C}{2v} symmetry.

The distinction becomes crucial in Faraday geometry. If the zero magnetic field response were taken as evidence of exact \sym{C}{2v} symmetry, the corresponding magnetic group, \sym{C}{2v}{}(\sym{C}{s}), would allow only two in-plane-polarised heavy hole trion transitions, in conflict with the four dominant transitions observed experimentally. Without any change in QD geometry, our calculations recover four dominant, nearly circularly polarised transitions, with an energetic ordering consistent with experimental spectra. The exact magnetic symmetry of the reduced-symmetry structure is therefore \sym{C}{s}{}(\sym{C}{1}), which permits all four transitions, while their dominant oscillator strengths and polarisations remain \sym{C}{3v}{}(\sym{C}{3})-like.

In conclusion, the same reduced-symmetry structure can therefore reproduce a \sym{C}{2v}-like optical response at zero field and a \sym{C}{3v}{}(\sym{C}{3})-like response in Faraday geometry, providing a unified description of both regimes without any restoration or enhancement of the underlying structural symmetry. Our work therefore clarifies the apparent discrepancy in the literature regarding the assigned symmetry of [111]-oriented arsenide-based QD systems in the presence and absence of magnetic fields. 

More generally, our results highlight the limitations of inferring QD symmetry from optical spectra alone. The absence of a transition does not demonstrate that it is symmetry forbidden, since a symmetry-allowed channel may simply carry an oscillator strength below the experimental detection limit. Consequently, a higher-symmetry-like spectrum does not uniquely establish the exact structural or magnetic symmetry. By contrast, the observation of a transition that is strictly forbidden by a candidate symmetry can exclude that symmetry. Symmetry assignments should therefore place greater weight on the presence of symmetry-excluding transitions than on the absence of weak optical features.

%% file: sections/5_Acknowledgements/acknowledgement.tex
\section*{Acknowledgements}

The authors wish to acknowledge the Irish Centre for High-End Computing (ICHEC) for the provision of computational facilities and support. N. D. O'C. thanks Dr. K. F. Karlsson for helpful discussions, and useful data for benchmarking. The authors thank Mr. J. McCloskey for help developing the CI methodology used in this work. This publication has emanated from research conducted with the financial support of Taighde Éireann - Research Ireland, formerly Science Foundation Ireland (SFI), under Grant Nos. 12/RC/2276 P2, 
15/IA/2864, 22/FFP-A/10930, 18/SIRG/5526, 22/FFP-P/11530, 23/RC/12219 and the European Innovation Council Pathfinder program under grant agreement No. 101185617 (QCEED).

%% file: Organisation/reset_SI.tex
\setcounter{page}{1}
\setcounter{section}{0}
\setcounter{equation}{0}
\setcounter{table}{0}

\gdef\thepage{S\arabic{page}}
\gdef\thesection{S\arabic{section}}
\gdef\theequation{S\arabic{equation}}
\gdef\thetable{S\arabic{table}}
\gdef\thefigure{S\arabic{figure}}

\section*{Supplemental Material}
\SIstartcontents
\SItableofcontents

%% file: appendices/numerics.tex
\section{\label{app:Num}Numerical Calculations}

This section provides further details of the numerical implementation underlying the 8-band \kp, self-consistent Hartree, and configuration-interaction (CI) calculations introduced in the main text. In particular, we describe the real-space discretisation, the [111]-rotated single-particle Hamiltonian, the treatment of magnetic fields within the finite-difference \kp framework, the material parameters, the Hartree procedure, the evaluation of Coulomb matrix elements, and the CI scheme.

\subsection{Real-Space Numerical Framework}
The strain field, piezoelectric potential, and single particle states are calculated in real space using finite differences, following established approaches for semiconductor heterostructures~\cite{StBi1997, WiLe2009, AnMo2008, EiVo2011, Ga2018, MiGa2018,Ca2013}. A uniform Cartesian grid with spacing $\Delta=0.25$~nm is used throughout, with Dirichlet boundary conditions imposed at the edge of the simulation domain. Spatially varying material parameters are assigned locally according to the composition of the QD, VQWr and surrounding matrix material.

\subsubsection{Single Particle State and Energy Calculations}
\input{appendices/numerics/sp}

\subsubsection{Strain Field}
\input{appendices/numerics/strain}

\subsubsection{Piezoelectric Potential}
\input{appendices/numerics/vpol}

\subsubsection{\label{app:sch}Self-Consistent Hartree Loop}
\input{appendices/numerics/sch}

\subsection{Multi-excitonic State Calculations}
The multi-excitonic states are calculated using a configuration interaction approach constructed from the self-consistent Hartree single particle states described above. Coulomb interactions between electron and hole states are included through direct and exchange matrix elements evaluated numerically on a real-space grid. The resulting many-body Hamiltonian is then diagonalised within the selected electron and hole basis to obtain the multi-excitonic states used for calculations of optical properties.

\subsubsection{Configuration Interaction Method}
\input{appendices/numerics/ci}

\subsubsection{Evaluation of Coulomb Matrix Elements}
\input{appendices/numerics/cmes}

%% file: appendices/numerics/sp.tex
The single particle electronic structure is evaluated using the 8-band \kp Hamiltonian of Ref.~\cite{Wi2003}, using the formulation of the Zeeman contribution from Ref.~\cite{MiGa2018}, in the total angular momentum basis $\ket{J,m_j}$. All terms are transformed consistently to the [111]-oriented coordinate system of Refs.~\cite{ScCa2011,MaO2014}, with axes,
\begin{align*}
    \hat{\mathbf{x}}&=\frac{1}{\sqrt{6}}[1,1,\bar{2}]\,,\quad
    \hat{\mathbf{y}}=\frac{1}{\sqrt{2}}[\bar{1},1,0]\,,\quad
    \hat{\mathbf{z}}=\frac{1}{\sqrt{3}}[1,1,1]\,.
\end{align*}
For completeness, the explicit Hamiltonian used after rotation to the coordinate system defined above is,
\begin{equation}
    H_{\mathbf{k}\cdot\mathbf{p}}^{[111]}=\begin{pmatrix}
        H_{6c6c} & H_{6c8v} & H_{6c7v} \\
        H_{6c8v}^\dagger & H_{8v8v} & H_{8v7v} \\
        H_{6c7v}^\dagger & H_{8v7v}^\dagger & H_{7v7v}
    \end{pmatrix}\,,
    \label{eq:kpHam}
\end{equation}
where labels $6c$, $8v$ and $7v$ denote the $\Gamma_{6}$ conduction, $\Gamma_{8}$ valence and $\Gamma_{7}$ split-off subspaces, respectively, and $\dagger$ denotes the adjoint. Although Eq.~\eqref{eq:kpHam} is written in $3\times3$ block form, the full Hamiltonian is $8\times8$. Each block is written as the sum of band edge, (0), kinetic, ($k$), bulk-inversion asymmetry, (BIA), strain, $(\varepsilon)$ and Zeeman, (Z), contributions:
\begin{equation}
    H_{ij} = H_{ij}^{(0)} + H_{ij}^{(k)} + H_{ij}^{(\mathrm{BIA})} + H_{ij}^{(\varepsilon)} + H_{ij}^{(\mathrm{Z})}\,,
\end{equation}
with $i,j\in\{6c, 8v, 7v\}$. Terms that do not contribute to a given block are omitted. The band edge term, $H_{ij}^{(0)}$, includes the electrostatic potential, $V_p$, which enters all diagonal blocks. The wavevector, $\mathbf{k}$, strain tensor, $\varepsilon_{ij}$, and the magnetic field, $\mathbf{B}$, are written in the [111] coordinate system defined above. The angular momentum matrices, $J_i$ (spin 3/2), the Pauli matrices, $\tau_i$, and the so-called cross space matrices, $T_i$ and $T_{ij}$ (with $U_i = T_i^\dagger$ and $U_{ij} = T_{ij}^\dagger$) are the conventional matrices tabulated in the literature~\cite{TrRo1979, Wi2003, LaLi2007}, such that the basis states $\ket{J, m_j}$ are quantised along the [111] direction.

The resulting blocks of the Hamiltonian are:
\input{appendices/numerics/Ham_Mat_Elements}
Each contribution is written as a term with an isotropic form, which is the same in any coordinate system, plus a correction arising from the cubic symmetry of the crystal, which carries the [111]-specific information. In the kinetic and strain terms, the anisotropy is collected in $\delta = \gamma_2 - \gamma_3$ and $\delta_b = b - \frac{d}{\sqrt3}$, both of which vanish for an isotropic crystal. The bulk-inversion asymmetry (BIA) term ($\propto C_k$) is entirely anisotropic. In the Zeeman contribution, the terms in $J_i^3$ are replaced by $\mathcal{Q}_i = J_i^3 - \tfrac{41}{20}J_i$ with $i\in\{x,y,z\}$, i.e.\ $J_i^3$ with its $J_i$-like (isotropic) component removed and absorbed into the isotropic $\kappa$ term. The remaining $q$ terms (in $\mathcal{Q}_i$) are purely anisotropic. Because the isotropic terms are unchanged by a relabelling of the axes, they are abbreviated with c.p., denoting the terms obtained by cyclic permutation $x\rightarrow y \rightarrow z \rightarrow x$ of the preceding term, while terms with anisotropy are written out in full. Throughout, $\{A,B\} = \tfrac{1}{2}(AB+BA)$.

The 8-band \kp Hamiltonian treats the coupling of the $\Gamma_{6c}$ conduction band to the $\Gamma_{8v}$ and $\Gamma_{7v}$ valence bands explicitly, through the Kane parameter, $P= \sqrt{\hbar^2E_p/2m_0}$, with $E_p$ given by Eq.\eqref{eq:ep_rescale}. The bulk parameters already contain this coupling and are therefore renormalised so that it is not counted twice. Following Ref.~\cite{Wi2003}:
\begin{align}
    \gamma_1 &= \gamma_1^{L} - \frac{E_p}{3E_g}\,,\\
    \gamma_{2,3} &= \gamma_{2,3}^{L} - \frac{E_p}{6E_g}\,,\\
    \kappa &= \kappa^L - \frac{E_p}{6E_g}\,,\\
    g' &= g + \frac{2E_p}{3}\left(\frac{1}{E_g}-\frac{1}{E_g+\Delta_{\text{so}}}\right)\,,\\
    A' &= \frac{\hbar^2}{2m_0}\left(\frac{1}{m^*}-\frac{2E_p}{3E_g} - \frac{E_p}{3(E_g+\Delta_{\text{so}})}\right)\,.
\end{align}
Where $\gamma_i^L$ and $\kappa^L$ are the Luttinger parameters, and $g$ and $m^*$ are the bulk electron $g$-factor and effective mass. These and the remaining parameters required for the calculations are listed in Table~\ref{tab:mat_param}. The envelope function equations are obtained using $k_\alpha \rightarrow -i \frac{\partial}{\partial \alpha}$, with $\alpha\in\{x,y,z\}$, and discretised on a Cartesian grid as mentioned above. In the above matrix elements, products of a spatially varying parameter, $F$, with second order wavevector components, written as $Fk_\alpha^2$ are expanded as $k_\alpha F k_\alpha = -\partial_\alpha(F\partial_\alpha)$, and are discretised via:
\begin{align}
        [\partial_\alpha(F\partial_\alpha\psi)]_j \approx \frac{F_{j+1}+F_j}{2\Delta^2}\psi_{j+1} - \frac{F_{j+1}+2F_j+F_{j-1}}{2\Delta^2}\psi_{j} + \frac{F_j+F_{j-1}}{2\Delta^2}\psi_{j-1}\,,
    \label{eq:d2dx}
\end{align}
where $j$ labels grid point number along the $\alpha$ direction. Terms containing $Fk_\alpha k_\beta$ ($\alpha\neq\beta$) are expanded in a symmetrised product~\cite{St2005}, $\tfrac{1}{2}(k_\alpha Fk_\beta + k_\beta F k_\alpha)$, and built from the centred first order derivative:
\begin{equation}
    [\partial_\alpha\psi]_j \approx \frac{\psi_{j+1} - \psi_{j-1}}{2\Delta}\,.
      \label{eq:ddxCentral}
\end{equation}
The spatially varying coefficient, $F$, is represented as a diagonal operator and the ordered products are formed by sparse matrix multiplication. The conduction-valence band coupling terms ($H_{6c8v}$, $H_{6c7v}$), contain terms linear in $k_\alpha$, e.g.\ $\sqrt3\,P\,T_\alpha\,k_\alpha$. For these terms we use the asymmetric ordering of Refs.~\cite{Fo1997, An2009}, $Pk_\alpha$ in the $\Gamma_6$ rows, and its adjoint $k_\alpha P$ in the $\Gamma_{8(7)}$ rows. The derivative is discretised with a fourth-order central difference in the interior and first order one sided differences at the two outermost points on each side of the simulation domain:
\begin{equation}
    [\partial_\alpha\psi]_j \approx 
    \begin{cases}
    \displaystyle \frac{\psi_{j+1} - \psi_{j}}{\Delta}, & j = 1,2 \\[5pt]
    \displaystyle \frac{\psi_{j} - \psi_{j-1}}{\Delta}, & j = n-1, n \\[5pt]
    \displaystyle \frac{-\psi_{j+2}+8\psi_{j+1}-8\psi_{j-1}+\psi_{j-2}}{12\Delta}, & 3\leq j\leq n-2\,.
    \end{cases}
    \label{eq:ddxOH4}
\end{equation}
Here $n$ is the number of grid points along the $\alpha$ direction. The adjoint blocks are formed as the Hermitian conjugate of the discretised operator, so the Hamiltonian matrix is exactly Hermitian.

The coupling to a magnetic field is introduced through the covariant wave vector operator~\cite{Lu1956a}:
\begin{equation}
    K_\alpha = -i\partial_\alpha + \frac{e}{\hbar}A_\alpha\,.
\end{equation}
Following Ref.~\cite{AnMo2008}, gauge invariance of the finite difference expansion is maintained by introducing the Peierls phase factors:
\begin{align}
    W_{j,j+s} &= \exp\left[\frac{ie}{\hbar}\int_{\mathbf{r}_j}^{\mathbf{r}_{j+s}}\mathbf{A}\cdot\mathrm{d}\boldsymbol{\ell}\right] = \exp\left[\frac{ie\,s\Delta}{2\hbar}\left(A_{\alpha,j} + A_{\alpha,j+s}\right)\right]\,,
    \label{Eq:MagPhase}
\end{align}
where $\mathbf{r}_{j+s} = \mathbf{r}_j+s\Delta\,\hat{\mathbf{e}}_{\alpha}$, and $s$ denotes the displacement in grid spacings along direction $\alpha$. Thus, $s=\pm1$ for nearest-neighbour terms and $s=\pm2$ where required for higher order finite difference stencils. Each neighbouring envelope function value, $\psi_{j+s}$, is multiplied by the corresponding phase factor, $W_{j,j+s}$, yielding a gauge invariant finite difference representation of the covariant derivatives.

\input{appendices/numerics/mat_params}

In the $\ket{J,m_j}$ basis used in our work, the Burt-Foreman ordering correction~\cite{Bu1992, Fo1997} is incorporated through anti-symmetrically ordered terms associated with the Zeeman contribution to the Hamiltonian~\cite{AnMo2008, GaGa2025a}, e.g., 
\begin{equation}
    \kappa B_z \longrightarrow \frac{i\hbar}{e}(K_x\kappa K_y - K_y\kappa K_x)\,.
    \label{eq:antisym}
\end{equation}
For spatially varying material parameters, $\kappa,\, g,\, q$, see Eq.~\eqref{eq:H6c6cZman} to Eq.~\eqref{eq:H7v7vZman}, these terms also generate interface contributions at zero magnetic field. To obtain single-particle states and energies, the eigenvalue problem related to the \kp Hamiltonian, Eq.~\eqref{eq:kpHam}, is solved using the Jacobi-Davidson method implemented in \textsc{SLEPc}~\cite{HeRo2005}.

The material parameters used in the calculations are listed in Table~\ref{tab:mat_param}. Here, the Kane energy, $E_p$, is chosen following Refs.~\cite{AnSe2010, VeSt2007}, to avoid a non-elliptic parameter set: 
\begin{equation}
    E_p = \left(\frac{1}{m^*}-1\right)\left(\frac{E_g(E_g+\Delta_{\text{so}})}{E_g + \frac{2}{3}\Delta_{\text{so}}}\right)\,.
    \label{eq:ep_rescale}
\end{equation}
Parameters for Al$_{x}$Ga$_{1-x}$As are obtained by interpolation between the binary compounds following Ref.~\cite{VuMe2001}:
\begin{equation}
    F(x) = xF^{\text{AlAs}} + (1-x)F^{\text{GaAs}} -x(1-x)C\,\, ,
\end{equation}
where $C$ is the bowing parameter. For all material parameters, $C=0$, except for the bandgap, $E_g$, where \mbox{$C = -0.127+1.310x$}.

%% file: appendices/numerics/Ham_Mat_Elements.tex
{\allowdisplaybreaks
\begin{align}
    H_{6c6c}^{(0)} &= \left[E_{\text{vb}} + E_{\text{g}} + V_p\right]\mathbf{1}_2\label{eq:H6c6cPE}\\
    H_{8v8v}^{(0)} &= \left[E_{\text{vb}} + V_p\right]\mathbf{1}_4\label{eq:H8v8vPE}\\
    H_{7v7v}^{(0)} &= \left[E_{\text{vb}} - \Delta_{\text{so}} + V_p\right]\mathbf{1}_2\label{eq:H7v7vPE}\\\nonumber\\
    H_{6c6c}^{(k)} &= \left[A'\mathbf{k}^2\right]\mathbf{1}_2\label{eq:H6c6c}\\
    H_{6c8v}^{(k)} &= \sqrt{3}\,P \left[k_x T_x + \mathrm{c.p.}\right] \label{eq:H6c8v}\\
    H_{6c7v}^{(k)}  &= -\frac{\sqrt{3}}{3}P\left[k_x\tau_x + \mathrm{c.p.}\right]\label{eq:H6c7v}\\
    H_{8v8v}^{(k)} &= \frac{\hbar^2}{2m_0}\Bigg[\left[-\gamma_1\mathbf{k}^2\right]\mathbf{1}_4 + \gamma_3\bigg[(k_x^2 -k_y^2)\left[J_x^2-J_y^2\right] \nonumber\\ &\qquad\qquad\qquad\qquad\qquad\quad+ (-k_x^2 -k_y^2 + 2k_z^2)\left[J_z^2-\frac{1}{3}J^2\right] + \left[4k_xk_y\left\{J_x,J_y\right\} + \mathrm{c.p.}\right]\bigg]\Bigg] \nonumber \\ 
     & \quad + \frac{\hbar^2}{2m_0}\delta\Bigg[\frac{1}{3}(k_x^2 - 2\sqrt2 k_xk_z - k_y^2)\left[J_x^2-J_y^2\right] + \frac{2}{3}\bigg[2k_y(\sqrt2 k_x + 2k_z)\left\{J_y,J_z\right\} \label{eq:H8v8v}\\ & \qquad\qquad\quad + (-\sqrt2k_x^2 + 4k_xk_z + \sqrt2 k_y^2)\left\{J_z,J_x\right\} + 2k_y(k_x + \sqrt2 k_z)\left\{J_x,J_y\right\}\bigg]\Bigg]\nonumber\\
    H_{8v7v}^{(k)} &= \frac{\hbar^2}{2m_0}\gamma_3\Bigg[\left(12k_xk_yU_{xy} + \mathrm{c.p.}\right) + 3\left((k_x^2-k_y^2)(U_{xx}-U_{yy}) + (-k_x^2-k_y^2+2k_z^2)(U_{zz})\right)\Bigg]\nonumber\\
    & \quad + \frac{\hbar^2}{2m_0}\delta\Bigg[4k_y(\sqrt2 k_x +2k_z)U_{yz} + 2(-\sqrt2 k_x^2 + 4k_xk_z + \sqrt2 k_y^2)U_{zx} \label{eq:H8v7v}\\
    & \qquad\qquad\quad + 4k_y(k_x + \sqrt2k_z)U_{xy} + (k_x^2 -2\sqrt2k_xk_z -k_y^2)(U_{xx}-U_{yy})\Bigg]\nonumber\\
    H_{7v7v}^{(k)} &= \left[-\frac{\hbar^2}{2m_0}\gamma_1\mathbf{k}^2\right]\mathbf{1}_2\label{eq:H7v7v}\\\nonumber\\
    H_{8v8v}^{(\mathrm{BIA})} &= C_k\Bigg[\left[\frac{5}{6}k_y\right]\mathcal{Q}_x -\left[\frac{5}{6}(k_x +\sqrt2k_z)\right]\mathcal{Q}_y + \left[\frac{5}{6}k_y\right]\left\{J_x, J_y^2-J_z^2\right\}\nonumber\\
    &\quad + \left[\frac{1}{6}(5k_x -3\sqrt2k_z)\right]\left\{J_y, J_z^2-J_x^2\right\} -\left[\frac{\sqrt2}{3}k_y\right]\left\{J_z, J_x^2-J_y^2\right\} \label{eq:H8v8vBIA}\\ &\quad - \left[\frac{\sqrt2}{3}k_x\right]\left(J_xJ_yJ_z + J_zJ_yJ_x\right)\Bigg]\nonumber\\
    H_{8v7v}^{(\mathrm{BIA})} &= iC_k \Bigg[k_yU_{yz} + k_xU_{zx} -\sqrt2k_yU_{xy} + \frac{\sqrt2}{2}k_x(U_{xx}-U_{yy})-\frac{3}{2}k_zU_{zz}\Bigg]\label{eq:H8v7vBIA}\\\nonumber\\
    H_{6c6c}^{(\varepsilon)} &= \left[a_c(\varepsilon_{xx}+\varepsilon_{yy}+\varepsilon_{zz})\right]\mathbf{1}_2\label{eq:H6c6cStr}\\
    H_{6c8v}^{(\varepsilon)} &= -2\sqrt3P\left(T_x\textstyle\sum_{j}\varepsilon_{xj}k_j + \mathrm{c.p.}\right)\label{eq:H6c8vStr}\\
    H_{6c7v}^{(\varepsilon)} &= \frac{2\sqrt3}{3}P\left(\tau_x\textstyle\sum_j\varepsilon_{xj}k_j + \mathrm{c.p.}\right)\label{eq:H6c7vStr}\\
    H_{8v8v}^{(\varepsilon)} &= \left[a_v(\varepsilon_{xx}+\varepsilon_{yy}+\varepsilon_{zz})\right]\mathbf{1}_4 + \sqrt3\,d\Bigg[\frac{1}{6}\bigg((-\varepsilon_{xx}+\varepsilon_{yy})(J_x^2-J_y^2) + (\varepsilon_{xx}+\varepsilon_{yy} - 2\varepsilon_{zz})(J_z^2-\frac{1}{3}J^2)\bigg)\nonumber \\ &\qquad\qquad\qquad\qquad\qquad\qquad\qquad\quad- \left(\frac{2}{3} \varepsilon_{xy}\left\{J_x,J_y\right\} + \mathrm{c.p.}\right)\Bigg]\label{eq:H8v8vStr}\\
    & \quad +\delta_b\Bigg[\frac{1}{6}(-\varepsilon_{xx} + 2\sqrt2\varepsilon_{xz} + \varepsilon_{yy})(J_x^2-J_y^2) -\frac{2}{3}(\sqrt2\varepsilon_{xy}+2\varepsilon_{yz})\{J_y,J_z\} \nonumber \\ 
    & \qquad\qquad + \frac{1}{3}(\sqrt2\varepsilon_{xx}-4\varepsilon_{xz}-\sqrt2\varepsilon_{yy})\{J_z,J_x\} - \frac{2}{3}(\varepsilon_{xy}+\sqrt2\varepsilon_{yz})\{J_x,J_y\}\Bigg]\nonumber\\
    H_{8v7v}^{(\varepsilon)} &= \sqrt3d\Bigg[\frac{1}{2}(-\varepsilon_{xx}+\varepsilon_{yy})(U_{xx}-U_{yy}) + \left(\frac{1}{2}\varepsilon_{xx} + \frac{1}{2}\varepsilon_{yy}-\varepsilon_{zz}\right)U_{zz} - 2\bigg(\varepsilon_{xy}U_{xy} + \mathrm{c.p.}\bigg)\Bigg]\nonumber\\
    &\qquad + \delta_b\Bigg[\left(-\frac{1}{2}\varepsilon_{xx} + \sqrt2\varepsilon_{xz} + \frac{1}{2}\varepsilon_{yy}\right)(U_{xx}-U_{yy}) -2(\sqrt2\varepsilon_{xy} + 2\varepsilon_{yz})U_{yz} \label{eq:H8v7vStr}\\
    & \qquad\qquad\quad+ (\sqrt{2}\varepsilon_{xx}-4\varepsilon_{xz}-\sqrt2\varepsilon_{yy})U_{zx} -2(\varepsilon_{xy}+\sqrt2\varepsilon_{yz})U_{xy}\Bigg]\nonumber\\
    H_{7v7v}^{(\varepsilon)} &= \left[a_v(\varepsilon_{xx}+\varepsilon_{yy}+\varepsilon_{zz})\right]\mathbf{1}_2\label{eq:H7v7vStr}\\\nonumber\\
    H_{6c6c}^{(\mathrm{Z})} &= \frac{1}{2}\mu_Bg'\left[B_x\tau_x + \mathrm{c.p.}\right]\label{eq:H6c6cZman}\\
    H_{8v8v}^{(\mathrm{Z})} &= -2\mu_B\Bigg[\left(\kappa+\frac{41}{20}q\right)B_xJ_x +\mathrm{c.p.}\Bigg] + \mu_Bq\Bigg[\frac{1}{6}(3B_x + 5\sqrt2B_z)\mathcal{Q}_x + \frac{1}{2}B_y\mathcal{Q}_y + \frac{4}{3}B_z\mathcal{Q}_z\nonumber\\ 
    &\qquad\qquad\qquad+\frac{1}{2}(B_x-\sqrt2B_z)\{J_x,J_y^2-J_z^2\} - \frac{1}{2}B_y\{J_y, J_z^2-J_x^2\} + \sqrt2B_x\{J_z,J_x^2-J_y^2\}\label{eq:H8v8vZman}\\
    & \qquad\qquad\qquad -\sqrt2B_y(J_xJ_yJ_z + J_zJ_yJ_x)\Bigg] \nonumber\\
    H_{8v7v}^{(\mathrm{Z})} &= -3\mu_B\Bigg[\left(1 + \kappa\right)B_xU_x + \mathrm{c.p.}\Bigg]\label{eq:H8v7vZman}\\
    H_{7v7v}^{(\mathrm{Z})} &= -\mu_B\bigg[\left(1 + 2\kappa\right)B_x\tau_x + \mathrm{c.p.}\bigg]\label{eq:H7v7vZman}
\end{align}
}

%% file: appendices/numerics/mat_params.tex
\begin{table}[htbp]
    \centering
    \begin{tabular}{|c|c|c|}
    \hline
        & AlAs & GaAs \\\hline\hline
       $E_{\text{vb}}\, [\text{eV}]$\textsuperscript{\cite{VuMe2001}} & -0.53 & 0 \\
       $E_{\text{g}}\, [\text{eV}]$\textsuperscript{\cite{VuMe2001}} & 3.099 & 1.519 \\
       $\Delta_{\text{so}}\, [\text{eV}]$\textsuperscript{\cite{VuMe2001}} & 0.28 & 0.341 \\
       $m^*$\textsuperscript{\cite{VuMe2001}} & 0.15 & 0.067 \\\hline\hline
       $\gamma_1^L$\textsuperscript{\cite{VuMe2001}} & 3.76 & 6.98 \\
       $\gamma_2^L$\textsuperscript{\cite{VuMe2001}} & 0.82 & 2.06 \\
       $\gamma_3^L$\textsuperscript{\cite{VuMe2001}} & 1.42 & 2.93 \\
       $\kappa^L$\textsuperscript{\cite{MlTa2005a}} & 0.12 & 1.28 \\
       $q$\textsuperscript{\cite{La1971}} & 0.03 & 0.04 \\
       $g$\textsuperscript{\cite{Wi2003}} & 1.52 & -0.44 \\
       $C_{\text{k}}\, [\text{eV}\mathring{\text{A}}]$\textsuperscript{\cite{Wi2003}} & 0.002 & -0.0034 \\\hline\hline
       $a_c\, [\text{eV}]$\textsuperscript{\cite{VuMe2001}} & -5.64 & -7.17 \\
       $a_v\, [\text{eV}]$\textsuperscript{\cite{VuMe2001}} & -2.47 & -1.16 \\
       $b_v\, [\text{eV}]$\textsuperscript{\cite{VuMe2001}} & -2.30 & -2.00 \\
       $d_v\, [\text{eV}]$\textsuperscript{\cite{VuMe2001}} & -3.40 & -4.80 \\\hline\hline
       $\epsilon_r$\textsuperscript{\cite{Wi2003}} & 10.06 & 12.4 \\
       $e_{14}\, [\text{C}/{m}^{2}]$\textsuperscript{\cite{CaSc2015}} & -0.055 & -0.205 \\
       $B_{114}\, [\text{C}/{m}^{2}]$\textsuperscript{\cite{CaSc2015}} & -1.61 & -0.99 \\
       $B_{124}\, [\text{C}/{m}^{2}]$\textsuperscript{\cite{CaSc2015}} & -2.59 & -3.21 \\
       $B_{156}\, [\text{C}/{m}^{2}]$\textsuperscript{\cite{CaSc2015}} & -1.32 & -1.28 \\\hline\hline
       $a_{\text{lat}}\, [\mathring{\text{A}}]$\textsuperscript{\cite{VuMe2001}} & 5.652 & 5.642 \\
       $C_{11}\, [\text{GPa}]$\textsuperscript{\cite{VuMe2001}} & 125.0 & 122.1 \\
       $C_{12}\, [\text{GPa}]$\textsuperscript{\cite{VuMe2001}} & 53.4 & 56.6 \\
       $C_{44}\, [\text{GPa}]$\textsuperscript{\cite{VuMe2001}} & 54.2 & 60.0 \\\hline
    \end{tabular}
    \caption{Material parameters underlying the \kp simulations. Band gap values at the $\Gamma$-point are considered here.}
    \label{tab:mat_param}
\end{table}

%% file: appendices/numerics/strain.tex
The strain field, arising from the lattice mismatch between GaAs and (Al,Ga)As, is calculated within continuum elasticity by minimising the elastic energy, $E_{el}$:
\begin{align}
    E_{el} &= \frac{1}{2}\int C_{ijkl}(\mathbf{r}) \varepsilon_{ij} \varepsilon_{kl}\,d^3r\,\, .
    \label{Eq:StrEnergy}  
\end{align}
Here $\mathbf{u}$ is the displacement field, $C_{ijkl}$ is the stiffness tensor and $\varepsilon_{ij}$ denotes the components of the strain tensor: 
\begin{align}
    \varepsilon_{ij} &= \frac{1}{2}\left(\partial_iu_j+\partial_j u_i\right) - \varepsilon^0_{ij}\,.
\end{align}
 The mismatch strain is defined as:
\begin{equation}
    \varepsilon^0_{ij}(\mathbf{r}) = \frac{a_0(\mathbf{r})-a_{\mathrm{ref}}}{a_{\mathrm{ref}}} \delta_{ij}\,,
\end{equation}
where $a_0(\mathbf{r})$ is the local unstrained lattice parameter. For the calculations presented here, $a_{\mathrm{ref}}$, the reference lattice constant, is taken as the VQWr material lattice constant. The stiffness tensor is expressed in the [111]-oriented coordinate system following Ref.~\cite{ScCa2011}.

The elastic energy of Eq.~\eqref{Eq:StrEnergy} is discretised by averaging over the eight combinations of forward and backward finite differences~\cite{StGr1999}; the resulting linear system is solved using conjugate gradients with algebraic multi-grid preconditioning in \textsc{PETSc}~\cite{BaGr1997}. The calculated elastic strain enters both the deformation potential terms of the 8-band \kp Hamiltonian and the first- and second-order piezoelectric polarisation.

%% file: appendices/numerics/vpol.tex
The strain tensor, $\epsilon_{ij}$, is used to calculate first- and second-order piezoelectric polarisation vector fields, $\mathbf{P}(\mathbf{r})$, in the [111]-oriented coordinate system above, using the derived expressions of Ref.~\cite{ScCa2011}. In the absence of free charges, the Poisson equation reads:
\begin{equation}
    \nabla\cdot\left[\epsilon_0\epsilon_r(\mathbf{r})\nabla\phi_p(\mathbf{r})\right] = -\rho_{\mathrm{pol}}(\mathbf{r})\,\, ,
    \label{eq:Poisson}
\end{equation}
where $\rho_{\mathrm{pol}} = -\nabla\cdot\mathbf{P}$ is the bound polarization charge density. The equation is discretised on the same finite-difference grid with Dirichlet boundary conditions and solved using \textsc{bicgstab} with block-Jacobi preconditioning in \textsc{PETSc}~\cite{BaGr1997}. The corresponding electron potential energy, $V_p = -e\phi_p$, where $e>0$ is the elementary charge, enters the single particle Hamiltonian as a grid point diagonal contribution.

%% file: appendices/numerics/sch.tex
The single particle basis is adapted to the exciton using a self-consistent Hartree procedure to improve convergence of subsequent many-body calculations~\cite{AbHa2006}. Starting from electron and hole ground states of $H_{\mathbf{k}\cdot\mathbf{p}}^{[111]}$, Eq.~(\ref{eq:kpHam}) the potentials connected to their respective charge densities are obtained at iteration $n$ from:
\begin{align}
    \nabla\cdot\left(\epsilon_0\epsilon_r\nabla\phi_e^{(n)}\right)&=-e|\psi_h^{(n)}|^2 \,\, ,\label{eq:elPotential}\\
    \nabla \cdot \left(\epsilon_0\epsilon_r\nabla\phi_h^{(n)}\right) &= +e|\psi_e^{(n)}|^2\,\,.\label{eq:hoPotential}
\end{align}
Here, $\phi_e$ and $\phi_h$ denote the potentials generated by the hole and electron charge densities, respectively. To obtain energy and state corrections for electrons in the presence of the hole, and vice versa, at each iteration step the eigenvalue problem:
\begin{equation}
    \left(H_{\mathbf{k}\cdot\mathbf{p}}^{[111]}+q_m\phi_m^{(n)}\right)\tilde{\psi}_{m,i}^{n+1} = \tilde{\varepsilon}_{m,i}\tilde{\psi}_{m,i}^{n+1}\,,
\end{equation}
is solved, with $m\in\{e,h\}$, $q_e = -e$ and $q_h = +e$. The Poisson equations, Eqs.~\eqref{eq:elPotential}, \eqref{eq:hoPotential}, use the numerical discretisation scheme and boundary conditions described above for Eq.~\eqref{eq:Poisson}. Iteration continues until changes in both the single particle eigenvalues and Hartree potential energies fall below $10^{-6}$~eV. The converged states, $\tilde{\psi}_{m,i}$, and eigenvalues, $\tilde{\varepsilon}_{m,i}$, form the basis for subsequent many-body CI calculations.

%% file: appendices/numerics/ci.tex
The Coulomb interaction between charge carriers confined in a QD is treated via the configuration interaction (CI) scheme using the self-consistent Hartree basis, $\tilde{\psi}$, introduced in Sec.~\ref{app:sch}~\cite{AbHa2006}. The many-body, CI Hamiltonian reads:
\begin{equation}
    H_{\mathrm{CI}}
    =\sum_{pq}t^e_{pq}e_p^\dagger e_q +\sum_{pq}t^h_{pq}h_p^\dagger h_q +H_{\mathrm{C}}\,\, ,
    \label{eq:CI_SP}
\end{equation}
where $t_{pq}^{e(h)}$ are the one body matrix elements in the Hartree basis. The Coulomb interaction operator $H_{\mathrm{C}}$, in general, contains the electron-electron, hole-hole and electron-hole direct and exchange contributions,
\begin{equation}
    H_C = \frac{1}{2}\sum_{ijkl}V^{hhhh}_{ijkl}h_i^\dagger h_j^\dagger h_k h_l + \frac{1}{2}\sum_{ijkl}V^{eeee}_{ijkl}e_i^\dagger e_j^\dagger e_k e_l - \sum_{ijkl}V^{ehhe}_{ijkl}e_i^\dagger h_j^\dagger h_k e_l + \sum_{ijkl}V^{eheh}_{ijkl}e_i^\dagger h_j^\dagger e_k h_l\, .
\end{equation}

The Hartree eigenenergies, $\tilde\varepsilon_{e, p}$ and $\tilde\varepsilon_{h,p}$, obtained from the self-consistent procedure of Sec.~\ref{app:sch}, already contain the mean field electron-hole interaction. To avoid double counting when the full Coulomb interaction is included in $H_{\mathrm{C}}$, this contribution is removed from the one-body matrix elements,
\begin{align}
    t^{e}_{pq} &= \tilde\varepsilon_{e,p} \delta_{pq} + e \langle \tilde\psi_{e,p}|\tilde\phi_e|\tilde\psi_{e,q}\rangle\label{eq:obmeel}\\
    t^{h}_{pq} &= \tilde\varepsilon_{h,p} \delta_{pq} - e\langle \tilde\psi_{h,p}|\tilde\phi_h|\tilde\psi_{h,q}\rangle\,.\label{eq:obmeho}
\end{align}
Because the Hartree states diagonalise the self-consistent Hartree Hamiltonian, the one body matrix elements of Eqs.~\eqref{eq:obmeel} and \eqref{eq:obmeho} are, in general, not diagonal. The above expressions recover the matrix elements of the original single particle Hamiltonian, in the Hartree relaxed basis. All Coulomb matrix elements entering $H_{\mathrm{C}}$ are evaluated using the same converged Hartree states.

%% file: appendices/numerics/cmes.tex
The Coulomb matrix elements (CMEs) entering the direct and exchange contributions to the many-body Hamiltonian are evaluated using the self-consistent Hartree states discussed in the previous section:
\begin{align}
    V^{abcd}_{ijkl} = \frac{e^2}{4\pi\epsilon_0}\iint\tilde\psi^{*}_{a,i}(\mathbf{r})\tilde\psi^{*}_{b,j}(\mathbf{r'}) \frac{1}{\epsilon_r(\mathbf{r},\mathbf{r'})\,|\mathbf{r}-\mathbf{r'}|}\, \tilde\psi_{c,k}(\mathbf{r'})\tilde\psi_{d,l}(\mathbf{r}) \, d\mathbf{r}\, d\mathbf{r'}\,,
    \label{eq:CME}
\end{align}
where $\{a,b,c,d\}\in\{e,h\}$. The dielectric screening is approximated by a spatially uniform relative permittivity, $\epsilon_{\mathrm{QD}}$, taken as that of the QD material. This approximation is appropriate for the small ($\approx8\%$) dielectric mismatch between GaAs and (Al,Ga)As. To numerically evaluate the CMES, following Ref.~\cite{LoRe2003}, we use:
\begin{align}
    G^{bc}_{jk} &= \mathcal{F}\Bigl[\tilde\psi^{*}_{b,j}(\mathbf{r'})\tilde\psi_{c,k}(\mathbf{r'})\Bigr]\mathcal{F}\Biggl[\frac{1}{|\mathbf{r}-\mathbf{r'}|}\Biggr]\,,\label{Eq:FFTConv}\\
    V^{abcd}_{ijkl} &= \frac{e^2}{4\pi\epsilon_0\epsilon_{\text{QD}}}\int \tilde\psi^{*}_{a,i}(\mathbf{r})\,
    \mathcal{F}^{\text{-1}}\bigl[G^{bc}_{jk}\bigr]\tilde\psi_{d,l}(\mathbf{r}) \, d\mathbf{r}\,.\label{eq:CMEFFT}
\end{align}
Where the Coulomb integral is evaluated numerically by expressing the inner integral as a convolution of the pair density with the Coulomb kernel, $1/|\mathbf{r}-\mathbf{r}'|$, and applying the convolution theorem~\cite{Br2000a} using Fast Fourier Transforms (FFTs, $\mathcal{F}$), Eqs.~\eqref{Eq:FFTConv}, \eqref{eq:CMEFFT}. This reduces the computational cost relative to direct evaluation of the 6-dimensional real-space integral.

%% file: appendices/GT.tex
\section{\label{app:GT}Derivation of Symmetry Allowed Selection Rules - Heavy-Hole Trion}

In this section we derive the positively charged trion, $X^+$, optical selection rules used in the main text for the magnetic point groups \sym{C}{2v}{}(\sym{C}{s}) and \sym{C}{s}{}(\sym{C}{1}). These reflect the \sym{C}{2v} reference symmetry and the exact \sym{C}{s} symmetry of the truncated QD structure considered in the main text, respectively. The magnetic field is applied in Faraday geometry, i.e., $\mathbf{B}\parallel [111]\parallel \hat{\mathbf{z}}$, which transforms like an axial vector~\cite{BrCr2010}.

In what follows, we use single arrows to denote the electron states:
\begin{align}
\ket{\uparrow} &= \left|\frac{1}{2},+\frac{1}{2}\right\rangle, \quad \ket{\downarrow} = \left|\frac{1}{2},-\frac{1}{2}\right\rangle\,,
\label{eq:electron_spins}
\end{align}
and, motivated by the predominantly HH character of the hole ground state obtained in the main text, double arrows for the pure-HH basis:
\begin{align}
    \ket{\Uparrow} &= \left|\frac{3}{2},+\frac{3}{2}\right\rangle\,, \quad \ket{\Downarrow}=\left|\frac{3}{2},-\frac{3}{2}\right\rangle\,.
    \label{eq:HH_spins}
\end{align}
For the $X^+$ ground state, the two holes occupy the same single particle orbital. The Pauli exclusion principle therefore requires the formation of an antisymmetric singlet state~\cite{BeKo2012},
\begin{equation}
    \ket{S_\text{HH}} = \frac{1}{\sqrt{2}}\left(\ket{\Uparrow\Downarrow} - \ket{\Downarrow\Uparrow} \right)\,.
    \label{eq:hh_singlet}
\end{equation}
The corresponding $X^+$ states are,
\begin{equation}
    \ket{X^+_i} = \ket{S_\text{HH}} \otimes \ket{e_i}\,,
    \label{AppEqGT:Trion_state}
\end{equation}
with $\ket{e_i}\in\{\ket{\uparrow},\ket{\downarrow}\}$, defined in Eq.~\eqref{eq:electron_spins}. Since the two hole singlet state is invariant under the symmetry operations considered below, the transformation properties of the $X^+$ states are determined by the electron state. As in the optical calculations described in the main text, recombination connects an initial trion state, $\ket{X^+_i} \in \{X^+_{\uparrow}, X^+_{\downarrow}\}$, to a final single hole state, $\ket{h_f} \in \{\Uparrow, \Downarrow\}$. The two initial electron configurations and two final HH states therefore give four possible trion transitions.

For a pure HH, the inter-band selection rules yield~\cite{BaGe2011}:
\begin{align}
    \ket{X_{\downarrow}^+} &\rightarrow \ket{\Downarrow}+ \sigma^+\,,\label{eq:HH_Rules_Sp}\\
    \ket{X_{\uparrow}^+} &\rightarrow \ket{\Uparrow}+ \sigma^-\,.
    \label{eq:HH_Rules_Sm}
\end{align}
Here the emitted photon polarisation follows from the recombination of a $\ket{\downarrow}$ electron with a $\ket{\Uparrow}$ hole, or a $\ket{\uparrow}$ electron with a $\ket{\Downarrow}$ hole. The opposite HH component of the singlet remains as the final hole state.

We now apply the \sym{C}{2v}{}(\sym{C}{s}) and \sym{C}{s}{}(\sym{C}{1}) magnetic symmetries to determine the dipole allowed selection rules.

\input{appendices/Group_Theory/C2v_Cs}

\input{appendices/Group_Theory/Cs_C1}

%% file: appendices/Group_Theory/C2v_Cs.tex
\subsection{\texorpdfstring{\sym{C}{2v}{}(\sym{C}{s})}{C2v (Cs)} Optical Selection Rules}

The higher symmetry zero-magnetic field reference is \sym{C}{2v} $= \{E,\sigma_{xy}, \sigma_{xz},\mathrm{C}_{\mathrm{2}x}\}$. For $\mathbf{B}\parallel \hat{\mathbf{z}}\parallel[111]$, the operations $E$ and $\sigma_{xy}$ leave the axial field invariant, whereas \sym{C}{2x} and $\sigma_{xz}$ reverse its sign and remain symmetries in combination with time reversal, $\mathcal{T}$, (anti-unitary). Following the notation in the main text, the corresponding magnetic point group is therefore:
\begin{equation}
    \mathrm{C}_{\mathrm{2v}}(\mathrm{C}_\mathrm{s}) = \{E, \sigma_{xy}\} + \mathcal{T}\{C_{2x}, \sigma_{xz}\}\,,
\end{equation}
with unitary subgroup $\mathrm{C}_\mathrm{s} = \{E, \sigma_{xy}\}$. Since $\sigma_{xy}$ remains a unitary symmetry operation at finite magnetic field, the corresponding single particle and trion states may be chosen as eigenstates of the mirror operator, $U(\sigma_{xy})$, and labelled by its eigenvalues.

\subsubsection{Electron and HH States}
We first determine the action of $\sigma_{xy}$, which is a unitary symmetry operation, on the electron and HH basis states introduced above. The horizontal mirror operation may be written as~\cite{BrCr2010}:
\begin{equation}
    \sigma_{xy} = \hat{P}\,\mathrm{C}_{2z}\,,
\end{equation}
where $\hat{P}$ is the spatial inversion operator. We denote by $U^{e}(g)$ and $U^{\mathrm{HH}}(g)$ the representations of a symmetry operation $g$ acting on the electron and HH states respectively. Since the $s$-like conduction band states are even under application of $\hat{P}$, $U^{e}(g)$ acts purely as a rotation~\cite{SaNa2020a}:
\begin{equation}
    U^{e}(\sigma_{xy})|\uparrow\rangle = -i|\uparrow\rangle\,,\quad
    U^{e}(\sigma_{xy})|\downarrow\rangle = +i|\downarrow\rangle\,.
\end{equation}
For the $p$-like HH states, which are odd under application of $\hat{P}$:
\begin{equation}
    U^{\mathrm{HH}}(\sigma_{xy})\ket{\Uparrow} = -i\ket{\Uparrow}\,, \quad
    U^{\mathrm{HH}}(\sigma_{xy})\ket{\Downarrow} = i\ket{\Downarrow}\,.
    \label{eq:HH_eigs}
\end{equation}
Thus, the electron and HH basis states carry the mirror eigenvalues $\pm i$. These complex eigenvalues reflect the spinor representation of the mirror operation in the corresponding unitary double group~\cite{BrCr2010}.

As noted above, the singlet state of Eq.~\eqref{eq:hh_singlet} is invariant under the symmetry operations considered here. Under $\sigma_{xy}$, this follows directly because its two constituent HH states carry opposite mirror eigenvalues, giving the singlet eigenvalue, $(+i)(-i) = +1$. The positively charged trion therefore inherits the mirror eigenvalue of the remaining electron:
\begin{equation}
    U^{X^+}(\sigma_{xy})|X^+_{\uparrow}\rangle =-i|X^+_{\uparrow}\rangle\,,\quad U^{X^+}(\sigma_{xy})|X^+_{\downarrow}\rangle =+i|X^+_{\downarrow}\rangle\,.
    \label{eq:trion_eigs}
\end{equation}

\subsubsection{Optical Selection Rules}

For the trion transitions defined above, the dipole matrix element is:
\begin{equation}
    M_{\alpha}^{fi} = \langle h_f|d_\alpha|X^+_i\rangle\, ,
    \label{eq:dipole_Mat_Elem}
\end{equation}
where $\alpha\in\{x,y,z\}$ denotes the light polarisation component. Since the dipole operator is proportional to the position operator, $\mathbf{d} \propto \mathbf{r}$, it transforms as a polar vector. The horizontal mirror acts as,
\begin{equation}
    (x,y,z)\xrightarrow{\sigma_{xy}}(x,y,-z)\,,
\end{equation}
so that the dipole components transform according to,
\begin{equation}
    d_x \rightarrow + d_x\,, \quad
    d_y \rightarrow + d_y\,, \quad
    d_z \rightarrow - d_z\,.
\end{equation}
We denote the mirror eigenvalues of the initial trion and final HH states introduced above, Eqs.~\eqref{eq:trion_eigs} and \eqref{eq:HH_eigs}, by $p_i$ and $p_f$, respectively, and $\chi_\alpha$ denotes the parity of the dipole component, $d_\alpha$, under $\sigma_{xy}$, with $\chi_x = \chi_y = +1$ and $\chi_z = -1$. Following the general matrix element selection rule of Bradley and Cracknell~\cite{BrCr2010}, a non-zero transition requires:
\begin{equation}
    \Gamma_f^*\,\otimes \Gamma_{d_\alpha} \otimes \Gamma_i \supset \Gamma_1\,. \label{eq:direct_product_cond}
\end{equation}
Since the relevant representations of the unitary \sym{C}{s} subgroup are one dimensional, each symmetry operation acts by multiplication with a single scalar rather than by mixing basis states. Under $\sigma_{xy}$, the condition of Eq.~\eqref{eq:direct_product_cond} therefore reduces to the product of the corresponding mirror eigenvalues. A symmetry allowed transition in this case requires:
\begin{equation}
    p_f^*\chi_\alpha p_i = 1\,. \label{eq:sym_allowed_cond}
\end{equation}

For the in-plane components of the dipole operator, $d_\alpha$, $\chi_x = \chi_y = +1$ so that Eq.~\eqref{eq:sym_allowed_cond} gives $p_f=p_i$. Using the mirror eigenvalues obtained above, the two symmetry allowed in-plane polarised transitions are therefore:
\begin{equation}
    \ket{X^+_\uparrow} \rightarrow \ket{\Uparrow} + \sigma^-\,,\quad \ket{X^+_\downarrow} \rightarrow \ket{\Downarrow} + \sigma^+\,.
\end{equation}

For the out of plane component $d_z$ of the dipole operator, $\chi_z = -1$, it follows that $p_f = -p_i$. The corresponding transitions are therefore:
\begin{equation}
    \ket{X^+_{\uparrow}} \rightarrow \ket{\Downarrow} + z\,,\quad \ket{X^+_{\downarrow}} \rightarrow \ket{\Uparrow} + z\,,
\end{equation}
entailing that these are symmetry allowed only in $z$-polarisation. Hence, \sym{C}{2v}{}(\sym{C}{s}) separates the four possible HH trion transitions into two circularly polarised in-plane transitions and two $z$-polarised transitions, as summarised in the main text.

%% file: appendices/Group_Theory/Cs_C1.tex
\subsection{\texorpdfstring{\sym{C}{s}{}(\sym{C}{1})}{Cs (C1)} Optical Selection Rules}

We now consider the exact \sym{C}{s}{}(\sym{C}{1}) symmetry of the truncated QD structure. At zero magnetic field, \sym{C}{s} $= \{E, \sigma_{xz}\}$. For $\mathbf{B}\parallel\hat{\mathbf{z}}\parallel[111]$, the vertical mirror, $\sigma_{xz}$, reverses the axial magnetic field:
\begin{equation}
    (0,0,B_z)\xrightarrow{\sigma_{xz}} (0,0,-B_z)\,,
\end{equation}
and therefore survives at fixed $B_z$ only in combination with time reversal. The finite field magnetic point group is thus:
\begin{equation}
    \mathrm{C}_\mathrm{s}(\mathrm{C}_1) = \{E,\mathcal{T}\sigma_{xz}\}\,,
\end{equation}
with the unitary subgroup, \sym{C}{1} $=\{E\}$. Unlike the \sym{C}{2v}{}(\sym{C}{s}) case considered above, there is therefore no non-trivial symmetry operation whose eigenvalues distinguish the finite field electron, hole or trion states. The mirror eigenvalue condition of Eq.~\eqref{eq:sym_allowed_cond} no longer determines the optical selection rules. The remaining constraints must instead be obtained from the remaining anti-unitary symmetry operation.

\subsubsection{Electron and HH States}

For the electron and HH spin states introduced above, Eqs.~\eqref{eq:electron_spins} and \eqref{eq:HH_spins}, the vertical mirror operation at zero field can be represented by:
\begin{equation}
    U^{e(\mathrm{HH})}(\sigma_{xz}) = -i\tau_y\,,
\end{equation}
where $\tau_y$ is the $y$ Pauli spin $1/2$ matrix~\cite{LaLi2007}. At zero magnetic field, its eigenstates form the two spinor irreducible representations, $\Gamma_3$ and $\Gamma_4$, with mirror eigenvalues $\pm i$ respectively. Since $\sigma_{xz}$ is no longer a unitary symmetry operation at finite magnetic field, $\mathbf{B}$, mixing between the two zero-field irreducible representations is symmetry allowed.

Using the standard time reversal representation, $\mathcal{T}=-i\tau_yK$~\cite{SaNa2020a}, where $K$ denotes complex conjugation, the remaining anti-unitary symmetry of the system may be written in the electron and HH spin bases as:
\begin{equation}
    \mathcal{A} = \mathcal{T}\,U^{e(\mathrm{HH})}(\sigma_{xz}) = -K\, .
    \label{eq:Antiunitary_SymOp}
\end{equation}
It follows that, $\mathcal{A}^2=+\mathbf{1}$, where $\mathbf{1}$ is the identity operator. Thus, unlike ordinary time reversal for a half integer spin state, $\mathcal{A}$ does not enforce a Kramers degenerate partner~\cite{BrCr2010}. For the current symmetry analysis, we denote by $H_\lambda(B_z)$, with $\lambda\in\{e, \mathrm{HH}\}$, the effective single particle Hamiltonian within the corresponding two spin subspace. Since $\mathcal{A}$ is a symmetry of both single particle Hamiltonians,
\begin{equation}
    \mathcal{A}\,H_\lambda(B_z)\,\mathcal{A}^{-1} = H_\lambda(B_z)\,.
\end{equation}
From Eq.~\eqref{eq:Antiunitary_SymOp}, it follows:
\begin{equation}
    H_\lambda^*(B_z) = H_\lambda(B_z)\,,
\end{equation}
because $K H_\lambda K = H_\lambda^*$. Each Hamiltonian, $H_{\lambda}$ can therefore be represented by a real matrix, and its non-degenerate eigenvectors can be chosen with real coefficients. We write,
\begin{align}
    \ket{e_i} &= a_i\ket{\uparrow} + b_i\ket{\downarrow}\,,\label{AppEqCsC1:ElectronState} \\
    \ket{h_f} &= c_f\ket{\Uparrow}+ d_f\ket{\Downarrow}\,,
    \label{AppEqCsC1:HoleState}
\end{align}
with $(a_i,\, b_i,\, c_f,\, d_f) \in\mathbb{R}$. These expressions retain only the electron and HH spin components relevant to the present selection rule analysis. The spatial dependence of the single particle wavefunctions is not written explicitly. Consequently, \sym{C}{s}{}(\sym{C}{1}) permits admixture of both basis components within each electron and HH state, while the magnitude of this mixing is determined by the calculated electronic structure, i.e., the geometry and composition of the dots, rather than by symmetry alone.

\subsubsection{Optical Selection Rules}

We first consider the in-plane optical response. Using the optical matrix element defined in Eq.~\eqref{eq:dipole_Mat_Elem}, we extend the polarisation index, $\alpha$ to the circular basis by defining,
\begin{equation}
    d_{\sigma^{\pm}} = \frac{d_x\mp id_y}{\sqrt2}\,.
\end{equation}
The corresponding circularly polarised matrix elements are denoted by $M_{\sigma^{\pm}}^{fi}$.

Using the HH recombination channels of Eqs.~\eqref{eq:HH_Rules_Sp} and \eqref{eq:HH_Rules_Sm}, together with the spin mixed electron and HH eigenstates of Eqs.~\eqref{AppEqCsC1:ElectronState} and \eqref{AppEqCsC1:HoleState}, gives
\begin{align}
    M_{\sigma^+}^{fi} &\propto b_id_f\,,\label{AppEqCsC1:sp_recomb}\\
    M_{\sigma^-}^{fi} &\propto -a_ic_f\,,\label{AppEqCsC1:sm_recomb}
\end{align}
where the relative minus sign follows from the antisymmetric singlet state of Eq.~\eqref{eq:hh_singlet}. Since none of the coefficients $a_i,\,b_i,\,c_f,\,d_f$ are required to vanish by \sym{C}{s}{}(\sym{C}{1}) symmetry, neither circular component is symmetry forbidden in general. Consequently, none of the four trion recombination channels is excluded by symmetry in either circular polarisation.

Using the inverse relations,
\begin{equation}
    d_x = \frac{d_{\sigma^+} + d_{\sigma^-}}{\sqrt{2}}\,, \qquad d_y = \frac{d_{\sigma^+} - d_{\sigma^-}}{\sqrt{2}}\,,
\end{equation}
the corresponding Cartesian matrix elements are,
\begin{equation}
    M^{fi}_x\propto (b_i d_f-a_i c_f)\,, \qquad M^{fi}_y\propto i(b_i d_f+a_i c_f)\,,
\label{EqAppCsC1:LinearPol}
\end{equation}
Since $a_i,\,b_i,\,c_f,\,d_f$ may be chosen real,
\begin{equation}
    M^{fi}_x \in \mathbb{R}\,, \qquad M^{fi}_y \in i\mathbb{R}\,,
\end{equation}
and therefore, the $x$- and $y$-polarised components are therefore exactly $90^\circ$ out of phase. Their magnitudes,
\begin{equation}
    |M_x^{fi}| \propto |b_id_f - a_ic_f|\,,\quad |M_y^{fi}| \propto |b_id_f + a_ic_f|\,,
\end{equation}
are not constrained to be equal. The in-plane response is therefore elliptically polarised in general, becoming circular only the special case $|M_x^{fi}|=|M_y^{fi}|$.

We finally consider emission polarised along the growth direction. The $z$-polarised dipole component is even under the zero-field mirror, $\sigma_{xz}$, since $z$ is unchanged by the transformation $(x,y,z)\rightarrow(x,-y,z)$. However, in a finite magnetic field, $\sigma_{xz}$ is not itself a unitary symmetry operation, and its mirror eigenvalues no longer provide restrictions on the dipole allowed transitions. Consequently, \sym{C}{s}{}(\sym{C}{1}) does not require the $z$-polarised matrix element of any of the four trion transitions to vanish.

%% file: appendices/OMEs.tex
\section{\label{App:NCOR}Analysis of Near-Circular Optical Response}

The exact \sym{C}{s}{}(\sym{C}{1}) magnetic symmetry permits a general elliptically polarised in-plane optical response. In principle, the reduced symmetry also permits field induced mixing of the electron and HH spin components, Eq.~\eqref{AppEqCsC1:ElectronState}, \eqref{AppEqCsC1:HoleState}. However, the calculated electron states remain almost completely spin pure along the [111] quantisation axis throughout the magnetic field range considered. The weak opposite helicity components obtained numerically cannot be attributed to appreciable electron spin mixing as a consequence.

The near-circular response instead follows from the symmetry constraints on the in-plane optical matrix elements. Denoting $x$- and $y$-polarised optical matrix elements of a transition between initial state $\ket{i}$ and final state $\ket{f}$ by $M_x^{fi}$ and $M_y^{fi}$, respectively, the higher symmetry \sym{C}{3v}{}(\sym{C}{3}) reference requires~\cite{DuGl2013}:
\begin{equation}
    M_y^{fi} = \pm iM_x^{fi}\,,
\end{equation}
so that the two components have equal magnitudes and are $90^\circ$ out of phase, corresponding to circular polarisation. For \sym{C}{s}{}(\sym{C}{1}), the surviving anti-unitary symmetry retains the $90^\circ$ phase relation, but does not require $|M_y^{fi}| = |M_x^{fi}|$. The in-plane response is therefore elliptically polarised in general.

Using,
\begin{equation}
    M_{\sigma^{\pm}}^{fi} = \frac{M_x^{fi}\mp iM_{y}^{fi}}{\sqrt{2}}\,,
\end{equation}
the $90^\circ$ phase relation causes the two in-plane amplitudes to interfere constructively in one circular component and destructively in the opposite component. The weak to strong circular intensity ratio may therefore be written as:
\begin{equation}
    R_{fi} = \left(\frac{I_{\text{weak}}^{fi}}{I_{\text{strong}}^{fi}}\right) \simeq \left(\frac{|M_x^{fi}|-|M_y^{fi}|}{|M_x^{fi}|+|M_y^{fi}|}\right)^2\,.
\end{equation}
where the assignment of the strong and weak helicities depends on the calculated optical matrix elements for a given transition.

In the limiting case $|M_x^{fi}|=|M_y^{fi}|$, the destructively interfering circular component vanishes and the transition is purely circular. For the four ground state trion transitions considered here, the calculated matrix elements satisfy $|M_x^{fi}|\simeq |M_y^{fi}|$, giving weak to strong ratios of approximately $R_{fi}\simeq 10^{-3}$. The calculated response is therefore only weakly elliptical despite the more general polarisation response permitted by the exact \sym{C}{s}{}(\sym{C}{1}) symmetry. This near-circular response reflects the optical matrix elements lying close to the more restrictive \sym{C}{3v}{}(\sym{C}{3}) limit and does not imply any restoration or elevation of the underlying structural symmetry.